\documentclass[sigplan,nonacm,screen]{acmart}

\usepackage{tikz}
\usepackage{amsmath}
\usepackage{graphicx}
\usepackage{graphics}
\usepackage{float}
\usepackage{subfigure}
\usepackage{multirow}
\usepackage{color}
\usepackage{tabularray}
\usepackage{kantlipsum}
\usepackage{soul}
\usepackage{utfsym}
\usepackage{booktabs}
\usepackage{enumitem}
\usepackage{url}

\newcommand*\whitecircled[1]{\tikz[baseline=(char.base)]{
            \node[shape=circle,fill=black,font=\bfseries,inner sep=0.3pt] (char) {\textcolor{white}{#1}};}}
\newcommand*\redcircled[1]{\tikz[baseline=(char.base)]{
            \node[shape=circle,fill=red,font=\bfseries,inner sep=0.3pt] (char) {\textcolor{white}{#1}};}}

\iftrue
\newcommand{\yi}[1]{{\color{blue}#1}}
\newcommand{\jie}[1]{{\color{red}[\textbf{\sc jie}: \textit{#1}]}}

\else
\newcommand{\yi}[1]{}
\newcommand{\jie}[1]{}
\fi

\definecolor{fcolor}{cmyk}{0.55,1,0,0.15}

\begin{document}
\begin{sloppypar}

\title[]{XBOF: A Cost-Efficient CXL JBOF with Inter-SSD Compute Resource Sharing}


\author{
\vspace{-2pt}
        {{Shushu Yi$^{\ast}$, Yuda An$^{\ast}$, Li Peng$^{\ast}$, Xiurui Pan$^{\ast}$, Qiao Li$^{\dagger}$, Jieming Yin$^{\ddagger}$, Guangyan Zhang$^{\S}$}}\\
\vspace{-1pt}
        {{Wenfei Wu$^{\ast}$}, Diyu Zhou$^{\ast}$, Zhenlin Wang$^{\P}$, Xiaolin Wang$^{\ast}$, Yingwei Luo$^{\ast}$, Ke Zhou$^{\sharp}$, Jie Zhang$^{\ast}$}\\
\vspace{-1pt}
        {\normalsize{Peking University$^{\ast}$, Mohamed bin Zayed University of Artificial Intelligence$^{\dagger}$}} \\
\vspace{-1pt}
       {\normalsize{Nanjing University of Posts and Telecommunications$^{\ddagger}$, Tsinghua University$^{\S}$}} \\
\vspace{-1pt}
       {\normalsize{Michigan Tech$^{\P}$, Huazhong University of Science and Technology$^{\sharp}$}}
}

\renewcommand{\shortauthors}{Shushu Yi et al.}

\begin{abstract}
Enterprise SSDs integrate numerous computing resources (e.g., ARM processor and onboard DRAM) to satisfy the ever-increasing performance requirements of I/O bursts. While these resources substantially elevate the monetary costs of SSDs, the sporadic nature of I/O bursts causes severe SSD resource underutilization in just a bunch of flash (JBOF) level.
Tackling this challenge, we propose \emph{XBOF}, a cost-efficient JBOF design, which only reserves moderate computing resources in SSDs at low monetary cost, while achieving demanded I/O performance through efficient inter-SSD resource sharing.
Specifically, XBOF first disaggregates SSD architecture into multiple disjoint parts based on their functionality, enabling fine-grained SSD internal resource management.
XBOF then employs a decentralized scheme to manage these disaggregated resources and harvests the computing resources of idle SSDs to assist busy SSDs in handling I/O bursts.
This idea is facilitated by the cache-coherent capability of Compute eXpress Link (CXL), with which the busy SSDs can directly utilize the harvested computing resources to accelerate metadata processing.
The evaluation results show that XBOF improves SSD resource utilization by 50.4\% and saves 19.0\% monetary costs with a negligible performance loss, compared to existing JBOF designs.
\end{abstract}

\maketitle

\section{Introduction}
\label{sec:intro}




I/O-intensive scenarios, such as cloud storage, large language model inference, and burst cache \cite{micheloni2013ssd,shi2017ssdup,mcallister2021kangaroo, liao2024adding,khasymski2012use,wang2020graphwalker}, eagerly demand extremely high I/O throughput to accelerate the ever-expanding dataset access.
Following this trend, solid-state drives (SSDs) have become one of the most indispensable storage media and have experienced continuous technical advancements in both \emph{scale-up} and \emph{scale-out} ways.

From the \emph{scale-up} aspect, SSD manufacturers integrate more hardware resources within each SSD device to enhance I/O parallelism.
For example, the emerging PCIe 5.0 SSDs \cite{phisone26} boost the performance of embedded ARM processors by 1.7$\times$ over the PCIe 4.0 ones \cite{phisone21} to accelerate the execution of SSD firmware. 
Moreover, SSDs typically equip large onboard DRAM (1 GB per TB flash \cite{samsung980pro}) to accommodate the entire metadata (mainly FTL mapping tables \cite{chung2009survey}) for fast access.
From the \emph{scale-out} aspect, SSD suppliers cluster tens of high-performance SSDs as \emph{just a bunch of flash} (JBOF) \cite{supermicro,sun2025scalio,jiang2025building,min2021gimbal,guo2023leed}, which aggregates the hardware resources from every SSD to deliver extremely high throughput.

\begin{figure}
 \centering
    \includegraphics[width=1\linewidth]{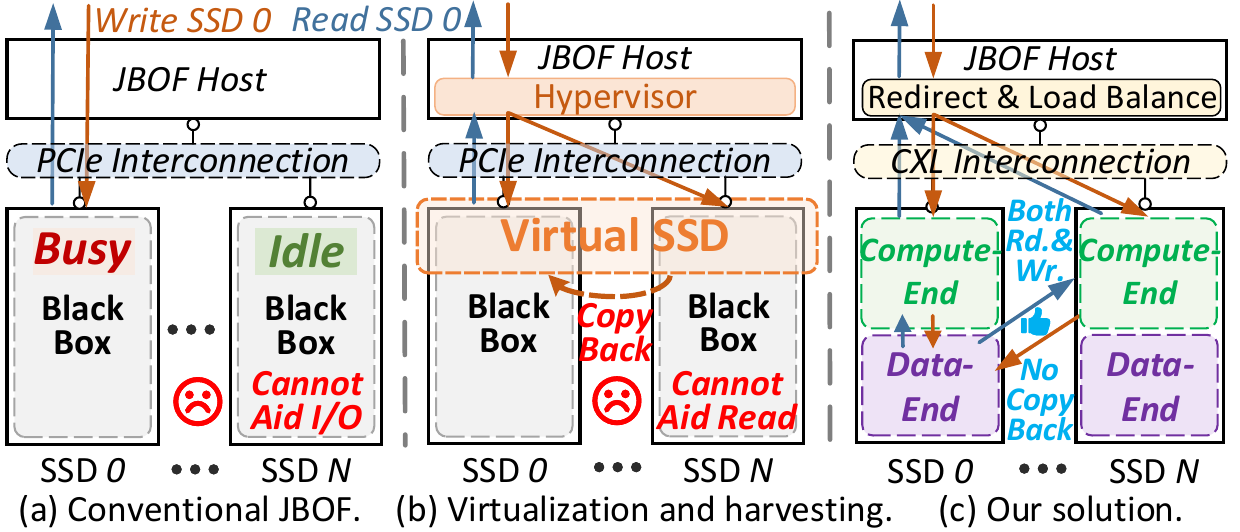}
    \vspace{-18pt}
    \caption{Comparison of different JBOF designs. \label{fig:intro}}
    \vspace{-10pt}
\end{figure}

Unfortunately, these technical trends place SSD consumers in a \emph{cost-utilization} dilemma.
To be more precise, while the increasing hardware resources elevate the \emph{bill of material (BOM)} costs of SSDs to satisfy the performance requirement of burst I/O, the sporadic nature of I/O bursts causes severe SSD resource underutilization in JBOF level (cf. Figure \ref{fig:intro}\textcolor{fcolor}{a}).
For instance, in 94.6\% uptime of a Tencent JBOF \cite{zhang2020osca} equipped with 25 drives, at least 20 drives are underutilized (i.e., bandwidth utilization is lower than 75\%, cf. $\S$ \ref{sec:bg_motivation}).
This is because in modern cloud platforms \cite{reidys2022blockflex,manvi2014resource}, storage drives (e.g., SSDs) are commonly allocated (or sold) to different tenants. These tenants utilize their own drives to serve different application instances with diverse I/O patterns, which experience I/O bursts at different times. 


%
A straightforward idea to improve utilization is storage virtualization and harvesting \cite{yang2018spdk,reidys2022blockflex,ambati2020providing,wang2021smartharvest,sun2025fleetio}. 
As shown in Figure \ref{fig:intro}\textcolor{fcolor}{b}, the hypervisor \cite{reidys2022blockflex,sun2025fleetio} can harvest idle SSDs by dynamically grouping a busy SSD with multiple idle ones as a \emph{virtual SSD}. Subsequently, parts of write requests originally targeting the busy SSD are redirected to the idle ones via the virtual SSD abstraction, leading to a higher burst performance and SSD utilization. 
Once the burst period concludes, these idle SSDs will be reclaimed and set aside for future harvesting. 
While this approach succeeds in exploiting idle SSDs, it unfortunately faces three prominent challenges:

\noindent $\bullet$ \emph{Coarse-grained}: 
Different I/O tasks impose varied degrees of burden on the computing (e.g., ARM processor and DRAM) and flash (e.g., flash channels) resources within SSDs. 
For example, 64 KB reads consume 95.4\% of processor clocks while only exploiting 42.2\% of flash times. 4 KB writes, in contrast, are flash-intensive (95.6\%) while keeping the processor underutilized (57.6\%, cf. $\S$ \ref{sec:ps_simple}). 
%
However, the hypervisor treats SSDs as monolithic black boxes, which leads to \emph{resource stranding} issues \cite{li2023pond,ambati2020providing,wang2021smartharvest}. For instance, when an SSD is flash-hectic for write bursts, its computing resources may still be idle. These computing resources cannot be harvested as the entire SSD is considered busy.

\noindent $\bullet$ \emph{Limited-profit}: 
The simple virtualization and harvesting approach yields minor benefits for read-dominated workloads.
This is because, without storing the target data of incoming read requests beforehand, the temporarily harvested SSDs cannot aid the busy SSDs in read services.
Our evaluation shows that this simple approach only brings 0.5\% throughput improvement in read-dominated workloads (cf. $\S$ \ref{sec:ps_simple}).  

%

\noindent $\bullet$ \emph{High-overhead}: 
Although redirecting write requests to the harvested SSDs can improve SSD utilization, this benefit comes at high costs. To be specific, when reclaiming the harvested SSDs, the hypervisor has to copy back the written data to the initial destination SSD for availability \cite{reidys2022blockflex}. Such write amplification drastically shortens SSD lifetimes (22.5\% reduction, cf. $\S$ \ref{sec:ps_simple}). 
Moreover, the centralized virtual SSD management in hypervisor can impose huge burdens \cite{kwon2020fvm,reidys2022blockflex} on the weak host CPU (e.g., 16 ARM cores in SuperMicro SSG-229J \cite{supermicro}) and compel it to become the performance bottleneck of JBOF (21.4\% throughput loss, cf. $\S$ \ref{sec:ps_simple}).



In response to these challenges, we introduce \emph{XBOF}, a cost-efficient JBOF design, which tackles the cost-utilization dilemma by only reserving moderate computing resources in SSDs while achieving satisfactory burst I/O performance through inter-SSD resource sharing (cf. Figure \ref{fig:intro}\textcolor{fcolor}{c}).
Our key insight is that the high-speed and cache-coherent \emph{Compute eXpress Link (CXL)} interconnections \cite{cxl,li2023pond,sun2023demystifying} can be harnessed to facilitate fine-grained, high-profit, and low-overhead SSD harvesting.  
Specifically, recognizing the black-box limitation of conventional SSDs, XBOF first disaggregates the SSD architecture into two parts, \emph{compute-end} and \emph{data-end}, based on their functionality.
Compute-end encloses computing resources, such as ARM processor and onboard DRAM, which are responsible for executing firmware tasks (e.g., address translation \cite{gupta2009dftl,chung2009survey}). 
Data-end comprises flash resources (e.g., flash channels) and data-related components (e.g., DMA engine), which are in charge of data transfer and flash I/O.
XBOF exposes them separately to the host and other SSDs via the high-speed CXL interconnection. This design enables fine-grained resource management and promises to mitigate the resource stranding issues. 

With the disaggregated SSD architecture, XBOF can benefit both read and write requests by harvesting idle compute-end to alleviate burdens of the SSDs, which are busy with firmware tasks. 
This idea is facilitated by the cache-coherent feature of CXL, with which the harvested compute-end can precisely operate the essential metadata of busy SSDs (e.g., FTL mapping table \cite{chung2009survey}) for I/O request handling.
Moreover, this design avoids detrimental copyback because it accelerates I/O requests by harvesting the stateless computing resource to expedite metadata processing while keeping the data path on the stateful flash memory unchanged.
%
Considering the overhead of centralized resource management, 
XBOF first leverages the global coherent memory constructed by CXL to enable inter-SSD communication. XBOF then implements a decentralized and self-governing resource management scheme in SSDs to relieve host CPU burdens.
Comprehensive evaluation results demonstrate that XBOF outperforms existing JBOF designs, improving SSD resource utilization by 50.4\%, saving 19.0\% BOM costs while having negligible performance loss.

Our main \textbf{contributions} can be summarized as follows: 

\begin{itemize}[leftmargin=10pt]

\item We deeply review the cost-utilization dilemma of JBOF and reveal that the black-box constraint of conventional SSDs can lead to severe resource stranding issues.

\item We propose a novel SSD architecture that disaggregates SSD internal resources into compute-end and data-end based on their functionality. This design lays the foundation for fine-grained and efficient SSD resource harvesting.

\item We propose XBOF, a cost-efficient JBOF design that reserves moderate computing resources in each SSD at low BOM costs while achieving demanded I/O performance by leveraging CXL to facilitate inter-SSD resource sharing.

\end{itemize}

\section{Background and Motivation}
\label{sec:background}



\subsection{JBOF and NVMe SSD}
\label{sec:bg_bg}

\begin{figure}
 \centering
    \includegraphics[width=1\linewidth]{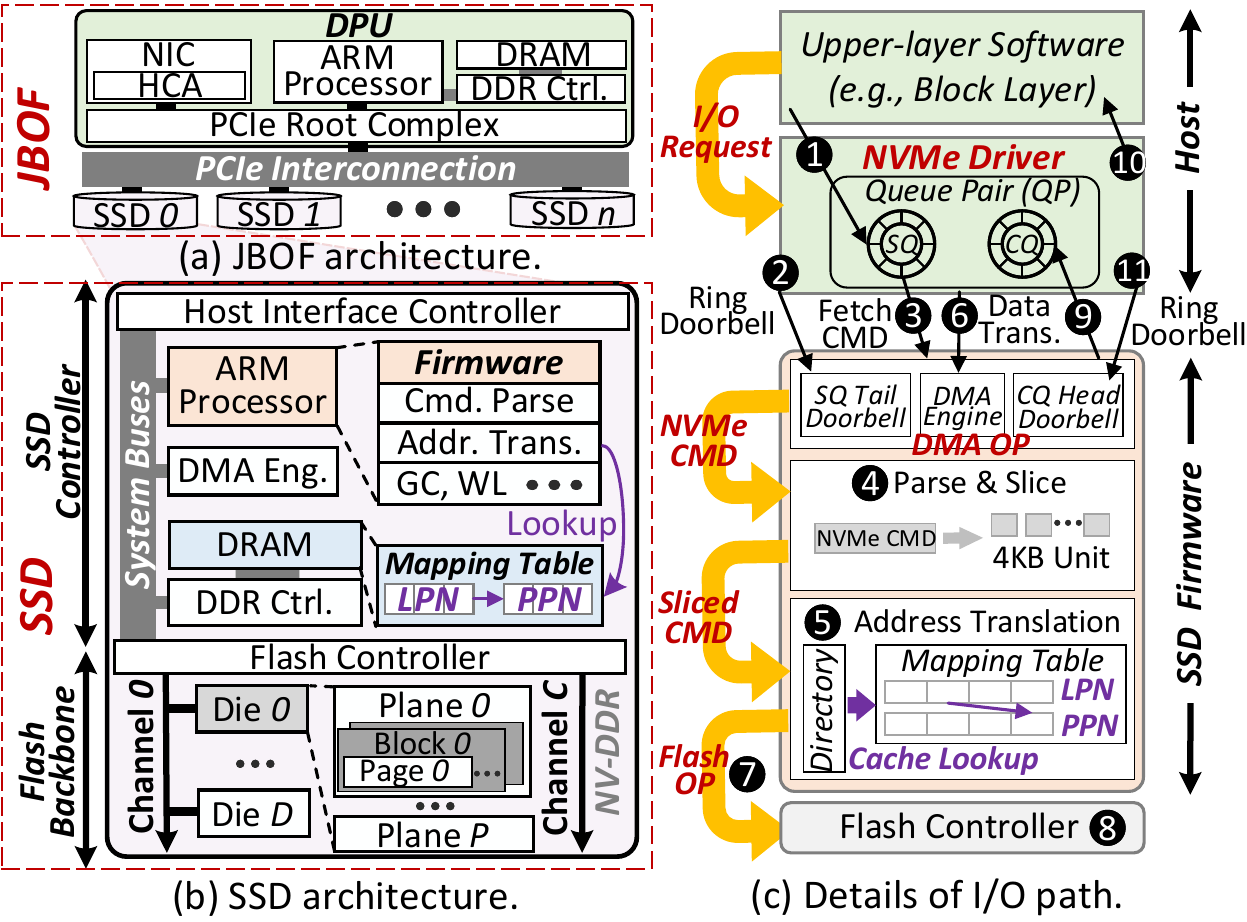}
    \vspace{-23pt}
    \caption{Details of the traditional JBOF system. \label{fig:background}}
    \vspace{-10pt}
\end{figure}

\noindent \textbf{JBOF architecture.}
Just a bunch of flash (JBOF) is a type of storage server that can incorporate multiple high-performance SSDs as a whole, thereby satisfying the ever-increasing performance demands in a scale-out manner.
As shown in Figure \ref{fig:background}\textcolor{fcolor}{a}, a JBOF typically comprises a few DPUs as the \emph{host} (or separated CPU, DRAM, and NIC in old-fashioned JBOF products \cite{sun2025scalio}).
The DPU has relatively wimpy computing power and connects to SSDs via PCIe interconnection.
For example, SuperMicro SSG-229J-5BU24JBF \cite{supermicro}, one of the up-to-date JBOF products, supports up to 24 SSDs with two NVIDIA BlueField-3 DPUs \cite{dpu}, each containing a 16-core ARM processor and 16 GB DRAM.
%


\begin{figure*}
    \begin{minipage}{0.595\textwidth}
        \includegraphics[width=1\linewidth]{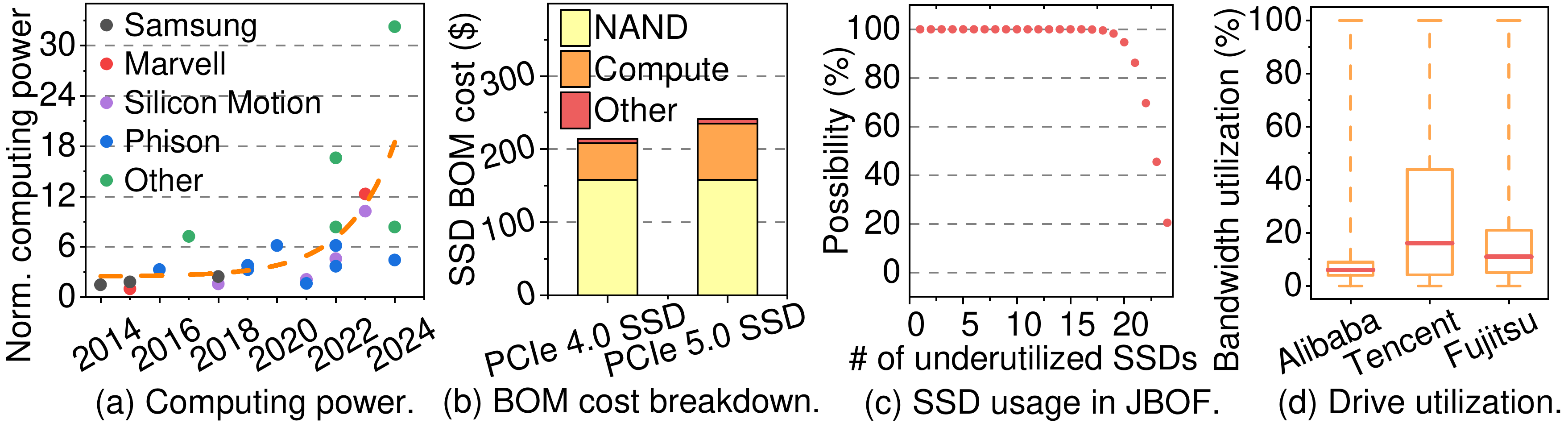}
        \vspace{-23pt}
        \caption{Analysis of the cost-utilization dilemma.\label{fig:costutilization}}
    \end{minipage}
    \begin{minipage}{0.395\textwidth}
        \includegraphics[width=1\linewidth]{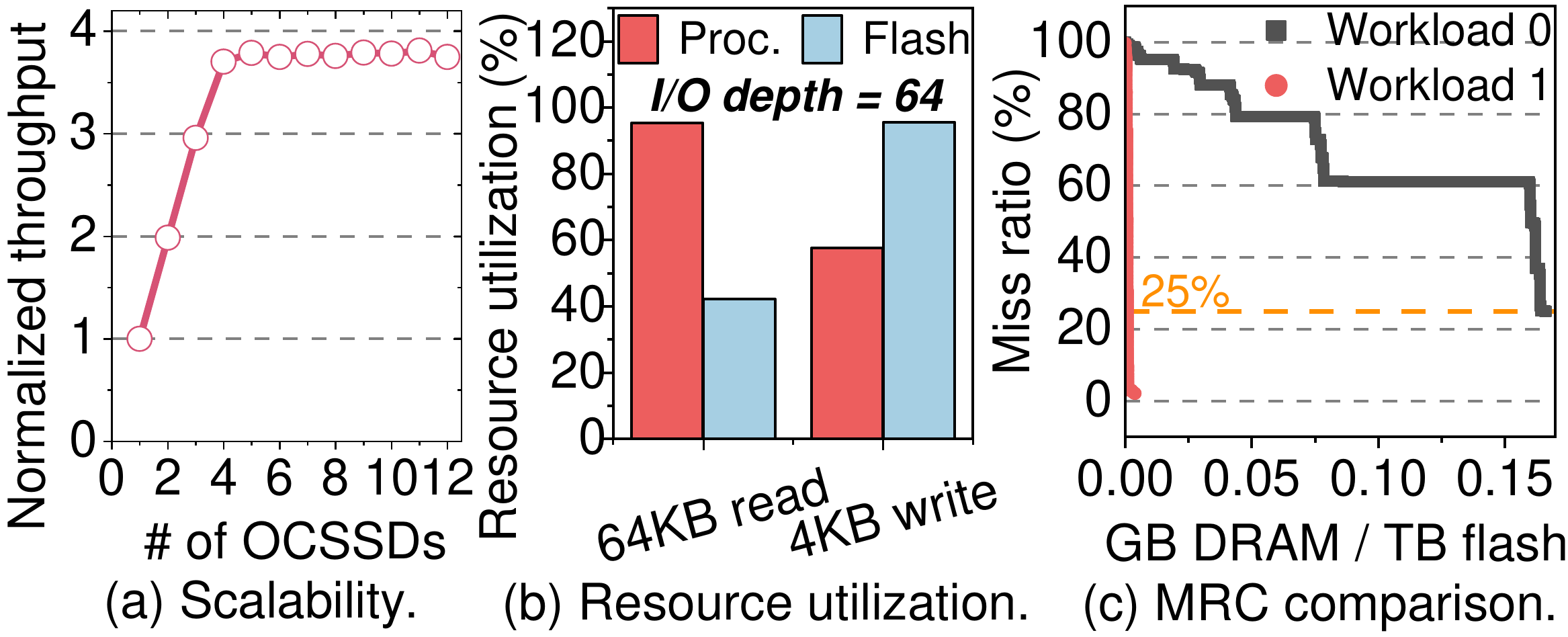}
        \vspace{-23pt}
        \caption{Preliminary study.\label{fig:preliminary}}
    \end{minipage}
\vspace{-7pt}
\end{figure*}

\noindent \textbf{SSD architecture.}
Figure \ref{fig:background}\textcolor{fcolor}{b} presents a typical architecture of modern SSDs \cite{PM1743}. 
The SSD controller connects to the host through PCIe lanes and a host interface controller. Moreover, it integrates an ARM processor, a DDR DRAM controller, and some specialized processing elements (e.g., DMA engine). 
The ARM processor is mainly responsible for performing firmware tasks (e.g., command parsing, address translation, and garbage collection). 
These components are coupled with a flash backbone through the flash controller. The flash backbone consists of 8 to 16 flash channels, each enclosing several flash dies. 
A flash die can be further divided into multiple flash planes, blocks, and pages.

\noindent \textbf{I/O path.}
Figure \ref{fig:background}\textcolor{fcolor}{c} illustrates the I/O path in JBOF systems.
When I/O requests arrive, the NVMe driver in the host submits NVMe commands to submission queues (SQ, \whitecircled{1}).
It then notifies the SSD of the command arrivals by ringing the SQ doorbells corresponding to the queues (\whitecircled{2}). 
Afterward, the SSD firmware operates the host interface controller to fetch NVMe commands from SQ (\whitecircled{3}).
SSD firmware then parses these commands, slices them into unit size (e.g., 4 KB, \whitecircled{4}), and translates the host \emph{logical address (LPN)} to the flash \emph{physical address (PPN)} by referring to the \emph{mapping table} in \emph{flash translation layer (FTL,} \whitecircled{5}\emph{)}.
The mapping table is persisted in the flash backbone for crash consistency and cached in onboard DRAM for high performance. A \emph{mapping directory} is used to locate cached mapping table entries.
SSD firmware also needs to orchestrate the host-SSD data transfers by issuing host \emph{DMA operations} to the DMA engine (\whitecircled{6}). 
Subsequently, the SSD firmware sends \emph{flash operations} to the flash controller (\whitecircled{7}), which performs flash I/O following the Open NAND Flash interface (ONFi) protocol \cite{ONFISpecs} (\whitecircled{8}). 
In the backward, the firmware writes the results to the completion queues (CQ) and then notifies the host by generating MSI-X interrupts \cite{dong2008sr} (\whitecircled{9}).
Finally, the NVMe driver reports request completions to upper-layer software (\whitecircled{\footnotesize{10}}) and acknowledges interrupts by updating the CQ doorbell (\whitecircled{\footnotesize{11}}).

\subsection{Motivation: Cost-Utilization Dilemma}
\label{sec:bg_motivation}

\noindent \textbf{High cost of SSD.} 
An obvious trend in SSD advancement is that SSD manufacturers tend to integrate numerous computing resources within SSDs to conduct firmware tasks rapidly, thereby boosting I/O performance. 
Figure \ref{fig:costutilization}\textcolor{fcolor}{a} presents the computing power of embedded processors in varied SSD controllers on Dhrystone v2.1 benchmark \cite{weicker1984dhrystone}. The results show that the computing power of embedded processors has increased exponentially over the last decade.
%
Moreover, enterprise SSDs \cite{PM1743,pcie6ssd} typically demand 1 GB onboard DRAM per TB flash capacity to accommodate their entire metadata (mainly the FTL mapping table) for fast access. 
These abundant computing resources cause high \emph{bill of material (BOM)} costs of SSDs. 
For instance, the computing resources (i.e., SSD controllers and DRAM) account for 23.2\% and 31.8\% of BOM costs to manufacture 4 TB PCIe 4.0 SSDs and PCIe 5.0 SSDs \cite{ChinaFlashMarket,dramexchange,TrendForce,Maxio} (cf. \texttt{Compute} in Figure \ref{fig:costutilization}\textcolor{fcolor}{b}).

\noindent \textbf{Low utilization of JBOF.}
Contrary to the continually increasing performance demands and BOM costs, SSD utilization in JBOF remains low due to the sporadic nature of I/O bursts.
Quantitative analysis reveals that in any uptime of a Tencent JBOF equipped with 25 drives \cite{zhang2020osca}, the probability of at least 20 drives being underutilized (i.e., under 75\% bandwidth utilization) is 94.6\% (cf. Figure \ref{fig:costutilization}\textcolor{fcolor}{c}). This is because in modern cloud platforms, especially the Infrastructure-as-a-Service (IaaS) cloud \cite{reidys2022blockflex,manvi2014resource}, storage drives are commonly allocated (or sold) to different tenants. These tenants utilize their own drives to serve varied applications with diverse I/O patterns,  which experience I/O bursts at different times.
This phenomenon exists in other JBOFs over diverse storage service providers. 
As depicted in Figure \ref{fig:costutilization}\textcolor{fcolor}{d}, the average drive bandwidth utilization is only 8.0\%, 27.8\%, and 15.3\% in Alibaba \cite{alibabaresource}, Tencent \cite{zhang2020osca}, and Fujitsu \cite{systor} clusters.

%
\yi{
}

\section{Preliminary Study}
\label{sec:preliminarystudy}
\subsection{Simple Solution and Its Challenges}
\label{sec:ps_simple}

The cost-utilization dilemma inspires future SSD and JBOF designs to be cost-efficient.
Ideally, future SSDs may only reserve moderate hardware resources at low BOM costs while achieving required I/O performance by harvesting underutilized resources within the same JBOF.

\noindent \textbf{JBOF consisted of open-channel SSDs.}
\emph{Open-channel SSD (OCSSD)} \cite{bjorling2017lightnvm,reidys2022blockflex,sun2025fleetio} is a representative SSD architecture that retains minimum hardware (e.g., flash controller and flash backbone) in SSD while relying on host computing resources and Linux \emph{LightNVM} driver \cite{linux_lightnvm} to conduct firmware tasks and cache metadata.
Although OCSSD is superior in low cost, it unfortunately hampers scalability and compatibility.
To be specific, the tens of OCSSDs in JBOF cause substantial computing overhead, compelling the wimpy JBOF DPU to become the performance bottleneck.
Figure \ref{fig:preliminary}\textcolor{fcolor}{a} shows the aggregated throughput of JBOFs with varying numbers of OCSSDs (cf. $\S$ \ref{sec:eval_setup} for experimental setups). 
Only 4 OCSSDs saturate the performance of OCSSD-based JBOF, because of the limited host resources.
%
%
Moreover, OCSSD faces severe compatibility issues as it requires huge manpower for OS and application adaptation to explicitly conduct firmware tasks, preventing it from wide deployment. 
As a result, Linux has removed LightNVM since v5.15 \cite{linux_515}. 
%

\noindent \textbf{SSD virtualization and harvesting.}
A more scalable and compatible solution is SSD virtualization \cite{yang2018spdk,russell2008virtio,min2021gimbal,dall2014kvm,bauman2015survey} and harvesting \cite{reidys2022blockflex,li2023pond,ambati2020providing,wang2021smartharvest,shan2018legoos,sun2025fleetio}. 
Specifically, the storage virtualization layer in hypervisor \cite{reidys2022blockflex, sun2025fleetio} can harvest the resources of idle SSDs by dynamically grouping a busy SSD (named \emph{borrower}) and several idle SSDs (named \emph{lender}) as a virtual SSD.
The virtual SSD is then exposed to users as a regular storage device. 
Thereafter, user write requests originally targeting the borrower can spread across both the borrower and lenders through the virtual SSD abstraction, thereby achieving higher burst I/O performance and SSD utilization.
Once the burst period concludes, these idle SSDs will be reclaimed and set aside for future harvesting.

\noindent \textbf{Challenges.}
While this approach succeeds in harvesting idle resources, it unfortunately faces prominent challenges:

\noindent $\bullet$ \emph{Coarse-grained SSD harvesting causes resource stranding issues \textbf{(Challenge 1)}}. 
We analyze the strain of different I/O tasks on SSD internal computing (i.e., ARM processor and DRAM) and flash (i.e., flash channels) resources, the results of which are shown in Figures \ref{fig:preliminary}\textcolor{fcolor}{b} and \ref{fig:preliminary}\textcolor{fcolor}{c}. 
64 KB sequential reads on an SSD with a 3-core 1 GHz ARM processor and 8-channel flash backbone (cf. $\S$ \ref{sec:eval_setup}) consume 95.4\% of the processor clocks while merely utilizing 42.2\% of flash clocks (cf. Figure \ref{fig:preliminary}\textcolor{fcolor}{b}).
In comparison, 4 KB sequential writes are flash-intensive (95.6\%), while leaving the processor underutilized (57.6\%).
Moreover, Figure \ref{fig:preliminary}\textcolor{fcolor}{c} illustrates the \emph{miss ratio curve (MRC)} \cite{shards,shakiba2024kosmo} of LRU-based metadata cache (i.e., FTL mapping table) in onboard DRAM.
\texttt{Workload 1} \cite{zhang2020osca} only consumes 0.001 GB DRAM (per TB flash) to achieve a 25\% miss ratio, while that is 0.17 GB for \texttt{Workload 0}.
In conclusion, different I/O tasks can impose varying levels of strain on computing and flash resources within SSDs.
Unfortunately, upper-layer software treats SSDs as monolithic black boxes, which leads to resource stranding issues. For example, when the flash backbone of an SSD is heavily engaged in 4 KB write bursts, its computing resources may remain idle. These idle resources cannot be lent to other SSDs in the simple approach because the entire SSD is considered busy.

%
\noindent $\bullet$ \emph{Limited profit in read-dominated workloads \textbf{(Challenge 2)}}. 
Our evaluation reveals that the simple virtualization and harvesting approach only brings 0.5\% and 0.8\% throughput improvements in Tencent \cite{zhang2020osca} and Alibaba \cite{alibabaresource,deng2018resource} workloads, respectively, where read requests dominate the I/O (cf. $\S$ \ref{sec:eval_benefit}). 
This is because the target data of read requests is exclusively stored in the borrower's flash backbone. Lenders cannot assist the borrower in serving read requests due to the lack of the target data in the lenders' flash backbone.

\noindent $\bullet$ \emph{High overheads brought by written data copyback \textbf{(Challenge 3.1)} and resource management \textbf{(Challenge 3.2)}}. 
%
In our tests, the simple approach causes 0.29 more drive writes per day (DWDP) on Tencent traces \cite{zhang2020osca}.
This is because, when lenders are reclaimed, the hypervisor must copy the written data back to the borrower for availability \cite{reidys2022blockflex}.
These extra writes lead to 22.5\% shorter SSD lifetime for enterprise SSDs \cite{PM1743} typically with 1 DWDP endurance.
Moreover, the centralized virtual SSD management in hypervisor can introduce substantial software overhead \cite{kwon2020fvm,reidys2022blockflex}, which compels the weak host CPU to become the performance bottleneck and causes significant throughput loss (e.g., 21.4\% in our tests on Tencent traces \cite{zhang2020osca}).
%

\subsection{Key Insight: Compute Express Link}
\label{sec:ps_cxl}

\noindent \textbf{CXL outline.}
\emph{Compute eXpress Link (CXL)} \cite{cxl} is an advanced interconnect standard designed to facilitate high-performance and cache-coherent communication among the host and various peripheral devices (e.g., SSD) \cite{jung2022hello,li2025bytefs,zhang2025skybyte}.
CXL comprises three sub-protocols: 
\emph{CXL.io} undertakes PCIe backward-compatible operations;
\emph{CXL.cache} empowers cache coherence in CXL fabric;
\emph{CXL.mem} enables device memory to be accessed by the host and other devices via \texttt{load/store} instructions.
For higher scalability, CXL 3.0 \cite{cxl3} introduces port-based routing and multi-level switching. These new features extend CXL fabric to rack-level and enable cache-coherent peer-to-peer communication among up to 4096 points (i.e., hosts or devices).
In this work, we conform to CXL 3.0 standard and equip SSDs with all three sub-protocols (i.e., as CXL Type-2 devices \cite{ji2024demystifying}), enabling cache-coherent memory access within the entire JBOF. 



%
\noindent \textbf{Opportunities.}
Recognizing the black-box constraint of conventional SSDs, we can first disaggregate the hardware resources of SSDs into multiple disjoint parts and then expose these parts separately to the host and other SSDs through the high-speed CXL interconnection. 
This design facilitates fine-grained resource management, laying a foundation to tackle the resource stranding issues (solution of \emph{\textbf{Challenge 1}}).
Additionally, with the support of cache-coherent memory access, 
the lender's processor can help the borrower handle I/O requests (e.g., command parsing and address translation) by directly operating the borrower's metadata stored in its onboard DRAM through CXL fabric. 
Moreover, the borrower can cache parts of its mapping table in the lender's DRAM for a lower miss ratio.
This method benefits both reads and writes and avoids data copyback overhead (solution of \emph{\textbf{Challenges 2 and 3.1}}), as it only harvests the stateless computing resources (i.e., processor and DRAM) to accelerate metadata processing, without redirecting data. 
%
Finally, we can exploit the cache-coherent memory to facilitate efficient inter-SSD communication. Thereafter, a self-governing and decentralized SSD resource management scheme can be implemented to alleviate the CPU burden imposed by the centralized virtualization and harvesting (solution of \emph{\textbf{Challenge 3.2}}).

\section{Design and Implementation}
\label{sec:design}

Inspired by the aforementioned preliminary analysis, we propose XBOF, a novel JBOF design that facilitates fine-grained, high-profit, and low-overhead inter-SSD resource sharing to tackle the cost-utilization dilemma. 

\subsection{Overview}
\label{sec:design_overview}

\begin{figure}
 \centering
    \includegraphics[width=1\linewidth]{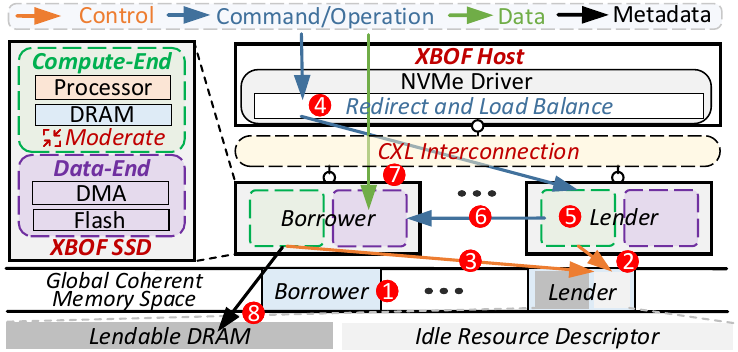}
    \vspace{-20pt}
    \caption{Overview of XBOF. \label{fig:overview}}
    \vspace{-10pt}
\end{figure}

\noindent \textbf{Base components.} 
Figure \ref{fig:overview} illustrates the overview of XBOF. Compared with existing JBOF designs, there are four main differences: 
(1) XBOF replaces conventional PCIe interconnections with CXL to enjoy its high performance and cache coherence;
(2) XBOF breaks the black-box constraint of traditional SSDs and enables fine-grained management of SSD internal resources ($\S$ \ref{sec:design_ssd});
(3) XBOF only reserves moderate computing resources (i.e., weaker processor and smaller DRAM) within SSDs at low BOM cost while satisfying burst I/O performance demands via resource harvesting;
(4) XBOF makes a minor modification to the host NVMe driver for I/O redirection and load balance.
For simplicity, we assume SSDs in XBOF are homogeneous (i.e., equipping the same hardware and running the same firmware), matching the common practice in JBOF markets \cite{supermicro}.
We will discuss heterogeneous scenarios in $\S$ \ref{sec:relatedwork}.

\noindent \textbf{Workflows.}
During device initialization (e.g., system reboot \cite{nvmebase}), each SSD exposes portions of its onboard DRAM via CXL interconnections. These exposed DRAM make up a global coherent memory space, facilitating inter-SSD communication (\redcircled{1}).
%
Afterward, if an SSD is idle (i.e., \emph{lender}), it calculates how many computing resources (i.e., processor and DRAM) can be lent out and announces this availability by writing its \emph{idle resource descriptors} (\redcircled{2}, $\S$ \ref{sec:design_management}).
When the computing resources in one SSD (i.e., \emph{borrower}) are in short supply (e.g., an I/O burst comes or a high DRAM miss ratio occurs), it searches all other SSDs' idle resource descriptors and chooses a lender for resource harvesting (\redcircled{3}, $\S$ \ref{sec:design_management}). 
This step can be repeated multiple times to borrow more resources from multiple SSDs.
For processor harvesting ($\S$ \ref{sec:design_processor}), the host NVMe driver redirects portions of NVMe I/O commands originally targeting the borrower to the lender (\redcircled{4}). 
The lender then assists in serving I/O commands with its idle processor by operating the borrower's metadata (\redcircled{5}). 
Subsequently, the lender sends DMA and flash operations to the DMA engine and flash controller of the borrower (\redcircled{6}).
The borrower then executes these operations to transfer data \textbf{directly} between the host and its flash backbone, without passing through the lender (\redcircled{7}).
%
For DRAM harvesting ($\S$ \ref{sec:design_dram}), the borrower can directly cache parts of its mapping table in the lender's DRAM (\redcircled{8}).
This harvested DRAM improves the cache hit ratio and avoids frequent flash access for the mapping table, leading to a higher I/O performance. 

\begin{figure}
 \centering
    \includegraphics[width=1\linewidth]{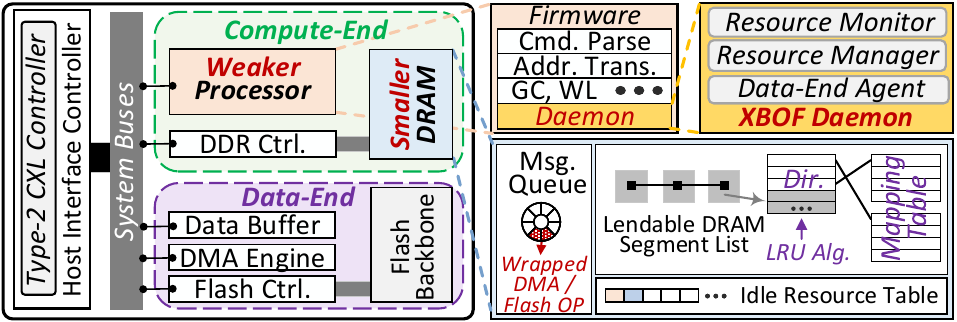}
    \vspace{-20pt}
    \caption{Disaggregated SSD architecture. \label{fig:disaggregate}}
    \vspace{-10pt}
\end{figure}

\subsection{Disaggregated SSD Architecture}
\label{sec:design_ssd}
Conventional SSDs are black boxes in which the onboard computing and flash resources are tightly coupled and invisible to external systems.
This agnostic causes resource stranding issues (cf. $\S$ \ref{sec:ps_simple}).
%
Tackling this challenge, XBOF employs a disaggregated SSD architecture that decouples SSD into two disjoint parts: \emph{compute-end} and \emph{data-end}, as shown in Figure \ref{fig:disaggregate}.  
Compute-end comprises computing resources, such as ARM processor, DDR controller, and onboard DRAM. 
It is responsible for executing SSD firmware tasks (e.g., I/O parsing and address translation).
Data-end encloses flash resources (e.g., flash controller and backbone) and data-related components (e.g., DMA engine and data buffer). 
This part is in charge of data transfer and flash I/O.
Moreover, a Type-2 CXL controller \cite{ji2024demystifying} is employed to facilitate CXL-related operations, such as exposing onboard DRAM to the host and peer SSDs in XBOF and operating the exposed DRAM of other SSDs coherently.
Specifically, during system initialization, the CXL controller of each XBOF SSD registers its local DRAM as \emph{global fabric-attached memory (G-FAM)} to the CXL fabric manager (FM) \cite{cxl3}.
Afterward, SSD processor can access peer SSDs' G-FAM via \texttt{load/store} instructions (e.g., \texttt{LDR/STR} in \texttt{aarch64} ISA \cite{armv8}), which will be interpreted as CXL \texttt{MemRd/MemWr} requests and then executed by the CXL controller.
Moreover, the CXL controller is responsible for maintaining the coherence of the SSD's local DRAM with \texttt{BISnp} and \texttt{BIRsp} requests (i.e., in HDM-DB mode \cite{cxl3}).

In the compute-end, an \emph{XBOF daemon} is implemented in SSD firmware and runs on the ARM processor.
It contains three components: 
(1) A \emph{resource monitor} is deployed to monitor the resource utilization of both compute-end and data-end. 
Specifically, for compute-end, the resource monitor periodically (e.g., 10 ms) polls the Performance Monitor Unit (PMU) \cite{contreras2005power,pmu} of the ARM processor to track its utilization (i.e., busy clocks).
DRAM resource is measured by the miss ratio of mapping table lookup.
For data-end, the resource monitor gets utilization from the embedded flash monitor module, which is implemented as hardware busy clock counters in the flash controller \cite{song2014cosmos,kwak2020cosmos+};
(2) A \emph{resource manager} is employed to borrow or lend computing resources based on the current utilization reported by the resource monitor;
(3) A \emph{data-end agent} bridges lender's compute-end with borrower's data-end. 
To be specific, the borrower's data-end agent maintains portions of onboard DRAM as globally visible message queues \cite{schuh2024cc}.
Thereby, the lender's compute-end can access the borrower's data-end by enqueueing wrapped DMA and flash operations (cf. $\S$ \ref{sec:bg_bg}) to the borrower's message queues.
The borrower's data-end agent then dequeues and unwraps these operations and sends them to the DMA engine and flash controller for data transfer and flash I/O.

\subsection{Decentralized Resource Management}
\label{sec:design_management}

\begin{figure}
 \centering
    \includegraphics[width=1\linewidth]{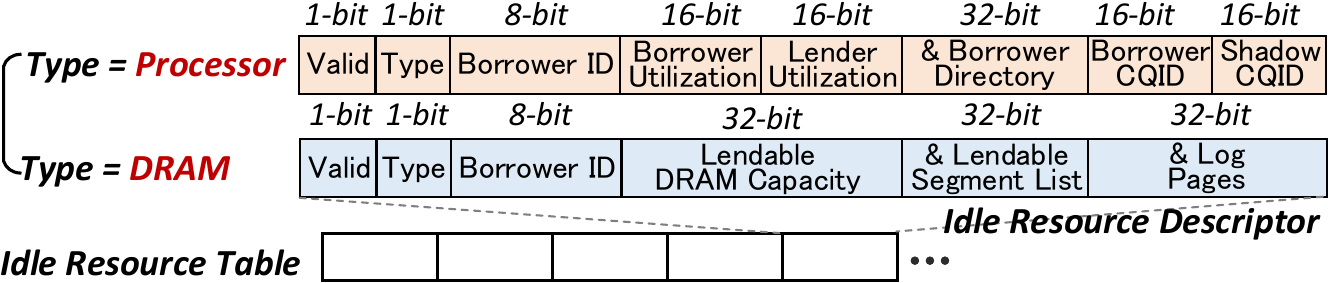}
    \vspace{-23pt}
    \caption{Data structures for resource management. \label{fig:descriptor}}
    \vspace{-10pt}
\end{figure}

As shown in Figure \ref{fig:descriptor}, each XBOF SSD maintains portions of its onboard DRAM as \emph{idle resource table}, consisting of multiple \emph{idle resource descriptors}. 
These data structures are visible to the host and all peer SSDs in XBOF and are synchronized with reader-writer locks \cite{calciu2013numa}.
%
There are two formats of idle resource descriptors used to describe idle processor and DRAM resources, respectively.
Both contain five messages:
(1) One valid bit points out whether this descriptor is valid;
(2) One type bit presents the type of idle computing resources (i.e., processor or DRAM);
(3) 8 bits record the identification of the borrower (0xFF means that the resource has not been borrowed). XBOF assigns a unique identification to each SSD during device initialization \cite{nvmebase};
%
(4) 32 bits depict the amount of idle resources. 
Specifically, for idle processor, these 32 bits are used to indicate the current processor utilization of both the borrower and the lender (16 bits for each) for load balance (cf. $\S$ \ref{sec:design_processor}).
Both the lender and the borrower will update this field periodically (e.g., 10 ms) after harvesting begins.
For idle DRAM, these 32 bits indicate the capacity of the lendable DRAM, maintained by the lender; 
(5) 64 bits record the essential information for resource sharing.
In particular, for idle processor, 32 bits indicate the address of the borrower's mapping table directory, while the other 32 bits record the CQIDs of borrower CQ and shadow CQ (16 bits for each) for I/O redirection (cf. $\S$ \ref{sec:design_processor});
For idle DRAM, 32 of the 64 bits point to the header of the lendable DRAM segment list, while the other 32 bits point to the start address of log pages for crash consistency (cf. $\S$ \ref{sec:design_dram}).

With the aforementioned data structures, XBOF enables decentralized resource management. Specifically, SSDs in XBOF can lend out their idle resources by writing the idle resource descriptors. Moreover, resource borrowing can be facilitated by searching the idle resource tables of all peer SSDs and choosing one (or more) idle SSD to harvest (i.e., atomically writing the borrower identification of the chosen idle resource descriptor). 
After harvesting begins, the lender, borrower, and host will periodically (e.g., 10 ms) check and update the idle resource descriptor to synchronize the status of harvesting.
Specifically, if the borrower no longer wants to borrow resources (e.g., the I/O burst is over), it sets the borrower identification of the idle resource descriptors to 0xFF.
Moreover, if the lender no longer wants to lend resources, it tags the valid bit of the descriptor as invalid.

\subsection{Transparent Processor Harvesting}
\label{sec:design_processor}


\noindent \textbf{Trigger conditions.}
XBOF SSD triggers processor resource harvesting based on the busy status of both the processor and data-end. 
They are regarded as busy if their current utilization exceeds a configurable watermark (e.g., 75\%), otherwise considered underutilized.
If both the processor and the data-end are busy, the SSD does nothing as it has no available processor for sharing. Also, borrowing extra processor yields minor as the data-end has been exhausted.
In comparison, whenever the processor is underutilized, the SSD can lend out this resource. This can happen when the SSD is bottlenecked by the flash backbone (e.g., write bursts, cf. $\S$ \ref{sec:ps_simple}) or the whole SSD is idle.
%
%
Lastly, if only the processor is busy (e.g., read bursts, cf. $\S$ \ref{sec:ps_simple}), the SSD can borrow processor resources to maximize the I/O parallelism of the data-end, achieving a higher throughput.
Correspondingly, SSDs cancel borrowing or lending when the status of their resources no longer satisfies the above trigger conditions. 

\begin{figure}
 \centering
    \includegraphics[width=1\linewidth]{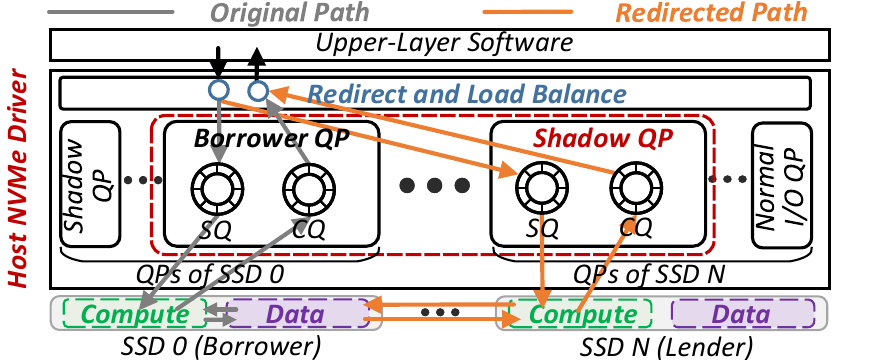}
    \vspace{-23pt}
    \caption{Transparent I/O redirection. \label{fig:shadow}}
    \vspace{-10pt}
\end{figure}

\noindent \textbf{Transparent I/O redirection.}
XBOF harvests idle processors by redirecting parts of the borrower's NVMe I/O commands to the lender. 
The lender then locates the borrower's mapping directory and table with the address recorded in the idle resource descriptor (cf. $\S$ \ref{sec:design_management}). 
Afterward, the lender can access the borrower's mapping tables and help serve I/O commands with its idle processor.
XBOF SSD employs reader-writer locks \cite{calciu2013numa} to enable efficient inter-SSD synchronization of the mapping table, inheriting from prior multi-core SSD designs \cite{zhang2020scalable,du2024pipessd}.
%
%
XBOF realizes I/O redirection by slightly modifying the host NVMe driver, which is the unique entrance of NVMe SSDs. 
%
%
This modification is transparent to the upper-layer applications (e.g., file system), ensuring compatibility.
As shown in Figure \ref{fig:shadow}, when initializing NVMe SSDs \cite{nvmebase}, XBOF reserves a few NVMe I/O queue pairs (QPs, each enclosing an SQ and a CQ) of each SSD as \emph{shadow QPs}. 
When lending processor resources, the lender points out the identification (e.g., CQID \cite{NVMeCommandSet}) of one of its shadow QPs in the idle resource descriptor.
Subsequently, the borrower chooses one of its normal I/O QPs (named \emph{borrower QPs}) for I/O redirection and also records its identification in the idle resource descriptor (cf. $\S$ \ref{sec:design_management}).
%
Thereafter, the host NVMe driver binds the borrower QP with the shadow QP and submits parts of NVMe I/O commands targeting the borrower SQ to the shadow SQ.
Following this, the lender fetches NVMe I/O commands from the shadow SQ and helps handle these commands.
In the backward, the NVMe driver collects results from both borrower CQ and shadow CQ and then commits I/O completions to upper-layer software. 
When ending harvesting, the shadow QP is unbound and waits for the next resource lending.

\noindent \textbf{Holistic load balance.}
The host NVMe driver selectively redirects I/O commands to the lender with a holistic load balance algorithm. 
This algorithm is two-fold. 
On the one hand, the NVMe weighted round-robin (WRR) feature \cite{nvmebase,woo2021d2fq} enables setting a weight for each NVMe I/O SQ, which indicates the priority of command fetching.
For example, if the weights of two SQs are 2 and 1, respectively, the SSD serves two I/O commands from the former SQ and then serves one command from the latter.
With this feature, XBOF can assign the shadow SQ of the lender with a low weight if it wants to minimize the performance impact on the lender's own I/O.
On the other hand, the host reads the idle resource descriptor periodically (e.g., 10 ms) to figure out the current processor utilization of both the borrower and lender (cf. $\S$ \ref{sec:design_management}). 
Then, it controls the number of NVMe I/O commands sent to the borrower and lender with the following formula:
$$
\frac{N_{borrow}}{N_{lend}} =\frac{U_{lend}}{U_{borrow}} \times \frac{\sum_{lend} W }{W_{shadowSQ}} \times \frac{W_{borrowSQ}}{\sum_{borrow} W } 
$$
$N_{borrow}$ and $N_{lend}$ represent the numbers of I/O commands sent to the borrower and lender, respectively.
$U_{borrow}$ and $U_{lend}$ are the processor utilization of the borrower and the lender.
$W_{borrowSQ}$ and $W_{shadowSQ}$ represent the weights of the borrower SQ and the shadow SQ. 
Lastly, $\sum_{borrow} W $ and $\sum_{lend} W $ are the total weights of all NVMe I/O SQs in the borrower and the lender.
With this formula, XBOF can balance processor utilization by selectively redirecting I/O commands. 
For example, if ${N_{borrow}}/{N_{lend}}$ is $3$, XBOF redirects I/O commands to the lender with a 25\% probability.

\subsection{Persistent DRAM Harvesting}
\label{sec:design_dram}
\noindent \textbf{Trigger conditions.} XBOF SSD manages DRAM resources in \emph{segments} (2 MB by default) and caches mapping table with LRU replacement algorithm. XBOF SSD decides whether to borrow or lend DRAM based on the miss ratio curve (MRC) of current mapping table lookup patterns.
Specifically, XBOF SSD adopts SHARDS \cite{shards}, a lightweight and efficient algorithm, to predict MRC online. 
Based on the predicted MRC, SSD can lend out all the spare DRAM segments, which has no help on a lower miss ratio (i.e., the cached mapping table will not be accessed in the near future), minimizing the effect of cache pollution from the borrower.
Correspondingly, SSD tries to borrow sufficient DRAM that reduces its miss ratio to below a given threshold (e.g., 10\%).

\noindent \textbf{Crash consistency.}
Borrower harvests DRAM by temporarily caching parts of its mapping table in the lender's lendable DRAM segments (i.e., as recorded in the idle resource descriptor, cf. $\S$ \ref{sec:design_management}).
While this idea can be facilitated by the cache-coherent capability of CXL, there still remains an open question in practice, that is, how to guarantee crash consistency of the \emph{offsite metadata} (i.e., the mapping table stored in the lender's DRAM).
To be specific, for high availability, enterprise SSDs are typically demanded to deliver \emph{power loss protection (PLP)} \cite{nvmebase}.
When an SSD suddenly loses power, the power hold-up circuit \cite{zheng2013understanding,narayanan2016ssd,tseng2011understanding} in the SSD immediately flushes the data and metadata buffered in the processor cache and onboard DRAM to the persistent flash backbone. 
This design ensures crash consistency of the SSD.
However, if the borrower's dirty mapping table is exclusively cached in lender's DRAM, the borrower cannot provide PLP to the offsite metadata. 
For instance, if the lender SSD is permanently unplugged from the JBOF, the borrower cannot recover its mapping table to locates data and suffers data loss. 

Tackling this issue, XBOF deploys a log-based crash consistency mechanism to protect the offsite metadata. 
%
In particular, when DRAM harvesting begins, the borrower vacates a 4 KB log page in its local DRAM for each harvested DRAM segment. 
Afterward, whenever modifying offsite metadata in the harvested segment, the SSD, either borrower or lender, needs to commit a log (e.g., redo log \cite{gray1981recovery}) to the log page associated with the segment.
Moreover, lenders need to ensure the log has been written back to the borrower with cacheline flush instructions (e.g., \texttt{DCCSW} for \texttt{aarch64} ISA \cite{DCCSW}).
%
When the log page of a harvested DRAM segment is full, the segment will be flushed back to the borrower's flash backbone, after which the corresponding log page can be cleared and reused.
Note that while the log operation introduces extra remote memory write overhead, it is relatively minor (i.e., hundreds of $ns$) compared to the flash I/O (i.e., tens of $\mu s$) caused by DRAM miss.
%
If the lender fails (e.g., multiple I/O timeouts occur \cite{nvmebase}), the host NVMe driver first notifies the borrower to recover its offsite metadata by replaying the logs in the local log pages.
Afterwards, NVMe driver resubmits in-flight NVMe I/O commands in the shadow SQ to the borrower and then unbinds the borrower QP with the shadow QP. 
If the borrower fails, the host NVMe driver first notifies the lender to recycle the lent resources by clearing the harvested DRAM segments and resetting the corresponding idle resource descriptors. Subsequently, NVMe driver clears the corresponding shadow QP, leaving it for future harvesting.

%

\subsection{Implementation}
\label{sec:design_imp}

\noindent \textbf{Prototype.}
We implement the host-side design of XBOF (e.g., I/O redirection and load balance, cf. $\S$ \ref{sec:design_processor}) in the NVMe driver of Linux kernel v5.15 \cite{linux_515} with 1 K LOC, following the NVMe specification \cite{nvmebase}.
Due to the lack of publicly available CXL 3.0 hardware, we prototype the firmware-side modification of our disaggregated SSD designs on DaisyPlus OpenSSD board \cite{kwak2020cosmos+,daisyplus}.
This board integrates a quad-core ARM Cortex-A53 processor, 2 GB DRAM, and adequate FPGA resources.
SSD firmware runs on the ARM processor, while the host interface controller and flash controller are implemented on the FPGA part.
We inherit the core functionalities (e.g., garbage collection) of the SSD firmware from DaisyPlus, but modify the I/O path with 2 K LOC (in C language) to realize XBOF daemon.
The data-end agent takes 114.2 ns, on average, to dequeue and unwrap a DMA/flash operation from the message queue (cf. $\S$ \ref{sec:design_ssd}).
Also, it takes 321.9 ns to commit a redo log to the local log pages for crash consistency (cf. $\S$ \ref{sec:design_dram}).
We cross-validate the performance model used in our simulator and emulator with these results.

\noindent \textbf{Simulator.}
We use SimpleSSD \cite{gouk2018amber,jung2017simplessd,simplessdgh}, a popular full-stack simulator, to evaluate XBOF.
It can accurately model the performance of host (e.g., CPU and DRAM) and modern SSD components (e.g., embedded ARM processor, DRAM, and flash backbone). 
We also extend SimpleSSD by 18 K LOC to support lots of detailed SSD techniques, such as incremental step pulse programming \cite{suh19953}, SLC cache \cite{yoo2020reinforcement,tripathy2022ssd}, and read retry \cite{ye2024achieving}.
These efforts ensure the accuracy of the simulator. 
%
To evaluate the CXL fabrics, we integrate ESF \cite{an2024novel}, a cycle-accurate CXL simulator, which can accurately model the features defined in CXL 3.0 standard \cite{cxl3}.
We use McPAT \cite{li2009mcpat} and DRAMPower \cite{chandrasekar2012drampower} to examine energy consumption.

\noindent \textbf{Emulator.}
For cross-validation, we also port XBOF to a NUMA-based emulation platform.
As recommended by prior work \cite{li2023pond,li2025bytefs,zhang2024omnicache,peng2025xharvest}, we mimic CXL fabric with cross-NUMA access and emulate each SSD with a dedicated socket (i.e., Intel Xeon 8562Y+ CPU \cite{intel8562Y}) running NVMeVirt \cite{kim2023nvmevirt}, a popular SSD emulator. 
However, constrained by the number of sockets (e.g., 2), such an emulation platform cannot accurately mimic the tens of SSDs in JBOF. Therefore, we opt to conduct most of our evaluation on the simulator, 
while taking a NUMA emulation in $\S$ \ref{sec:eval_numa}.

\section{Evaluation}
\label{sec:evaluation}
\subsection{Experimental Setup}
\label{sec:eval_setup}

\noindent \textbf{System configurations.} 
As listed in Table \ref{tab:config}, we configure the simulated JBOF system based on SuperMicro SSG-229J-5BU24JBF \cite{supermicro}, one of the up-to-date JBOF products, which supports at most 12 SSDs per DPU.
%
%
The simulated DPU contains a 16-core 2.1 GHz ARM processor and 16 GB DRAM, aligning with BlueField-3 \cite{dpu}, and acts as JBOF host (cf. $\S$ \ref{sec:bg_bg}).
The simulated SSD follows the configuration of commodity storage device \cite{micron9550,PM1743,crucialt705}, which delivers 14 GB/s and 10 GB/s peak read and write bandwidths. 
The SSD's processor encloses 6 ARM cores with \texttt{aarch64} ISA \cite{armv8} running at 1 GHz frequency.
In addition, its onboard DRAM can accommodate entire mapping table (i.e., 1 GB per TB flash).
%
%
%

\noindent \textbf{JBOF platforms.}
We compare XBOF with six other designs.
(1) \texttt{Conv}: conventional JBOF design, in which all SSDs equip abundant computing resources (i.e., 6 ARM cores and 1 GB DRAM per TB flash) for high I/O parallelism;
(2) \texttt{OC}: OCSSD-based JBOF design that reserves minimum computing resources in SSDs but utilizes host resources to execute firmware and cache metadata (cf. $\S$ \ref{sec:ps_simple});
(3) \texttt{Shrunk}: compared to \texttt{Conv}, it \textbf{shrinks} the computing resources in each SSD. By default, it halves the resources (i.e., 3 ARM cores and 0.5 GB DRAM per TB flash). We also evaluate different reservation settings in $\S$ \ref{sec:eval_sensitivity};
(4) \texttt{VH}: based on \texttt{Shrunk}, it uses the simple virtualization and harvesting approach (cf. $\S$ \ref{sec:ps_simple}) to improve I/O performance; 
(5) \texttt{VH(ideal)}: an ideal variant of \texttt{VH}, in which no copyback is required;
(6) \texttt{ProcH}: \texttt{Shrunk} with our processor harvesting designs (cf. $\S$ \ref{sec:design_processor});
(7) \texttt{XBOF}: a cost-efficient JBOF design that includes all the techniques proposed in this paper. It reserves the same amount of computing resources in each SSD as \texttt{Shrunk}.

\begin{table}
\centering
\resizebox{\linewidth}{!}{%
\setlength{\tabcolsep}{1pt}
\begin{tabular}{|c|ccc|}
\hline
{\textbf{Host}} & \multicolumn{3}{c|}{16-core 2.1 GHz ARM processor and 16 GB DDR5-5600 DRAM} \\ \hline
\multirow{6}{*}{\textbf{\begin{tabular}[c]{@{}c@{}}\\ \\ SSD\end{tabular}}} & \multicolumn{1}{c|}{\textbf{Performance}} & \multicolumn{1}{c|}{\textbf{ARM processor}} & \textbf{Flash backbone} \\ \cline{2-4} 
 & \multicolumn{1}{c|}{\begin{tabular}[c]{@{}c@{}}Read/Write: \\ 14/10 GB/s\end{tabular}} & \multicolumn{1}{c|}{\begin{tabular}[c]{@{}c@{}}6 Cores @ 1 GHz \\ ISA: aarch64 (ARMv8)\end{tabular}} & \begin{tabular}[c]{@{}c@{}}8 Channel (2400 MT/s, 8-bit) /\\ 8 Die / 4 Plane / 1024 Block /\\  1024 Page / 16 KB, 4TB in total\end{tabular} \\ \cline{2-4} 
 & \multicolumn{1}{c|}{\textbf{Detailed techniques}} & \multicolumn{1}{c|}{\textbf{DRAM}} & \multirow{4}{*}{\begin{tabular}[c]{@{}c@{}}Read/Program: LSB: 30/200 us, \\ CSB: 45/280 us, \\ MSB: 60/400 ms.\\ Erase: 3 ms\end{tabular}} \\ \cline{2-3}
 & \multicolumn{1}{c|}{\multirow{3}{*}{\begin{tabular}[c]{@{}c@{}}ISPP, multi-plane,\\ SLC cache, read retry...\end{tabular}}} & \multicolumn{1}{c|}{\multirow{3}{*}{\begin{tabular}[c]{@{}c@{}}1 GB per TB flash,\\ DDR4-3200\end{tabular}}} &  \\
 & \multicolumn{1}{c|}{} & \multicolumn{1}{c|}{} &  \\
 & \multicolumn{1}{c|}{} & \multicolumn{1}{c|}{} &  \\ \hline
\textbf{CXL} & \multicolumn{3}{c|}{CXL 3.0 / PCIe 6.0 * 2 lanes = 16 GB/s per SSD, 256B FILT, tree topology} \\ \hline
\multicolumn{1}{|l|}{\multirow{2}{*}{\textbf{\begin{tabular}[c]{@{}l@{}}Energy\\param.\end{tabular}}}} & \multicolumn{3}{c|}{\multirow{2}{*}{\begin{tabular}[c]{@{}c@{}}Flash op. voltage=3.3V, $I_{READ, PROG, ERASE}$=25mA, $I_{BUSIDLE}$=5mA, $I_{STDBY}$=10uA \\$CXL/PCIe_{PHY}$=6 pJ/bit; SSD processor=6.45W; DRAM read/write=22 pJ/bit\end{tabular}}} \\
\multicolumn{1}{|l|}{} & \multicolumn{3}{c|}{} \\ \hline
\end{tabular}
}
\vspace{-5pt}
\caption{System configurations.\label{tab:config}}
\vspace{-25pt}
\end{table}

\begin{table}
\centering
\resizebox{\linewidth}{!}{%
\setlength{\tabcolsep}{2pt}
\begin{tabular}{|c|c|c|c|c|c|c|c|}
\hline
\textbf{Workloads} & \textbf{src}    & \textbf{DAP}       & \textbf{MSNFS}     & \textbf{mds}       & \textbf{YCSB-A} & \textbf{Fuji-0} & \textbf{Fuji-1} \\ \hline
\textbf{Read ratio (\%)}      & 11.3 & 56.2 & 67.2  & 92.8 & 98.0  & 82.7  & 86.3 \\ \hline
\textbf{Avg. read size (KB)}  & 8.1  & 62.1 & 9.6   & 60.1 & 9.5   & 35.7  & 32.7 \\ \hline
\textbf{Avg. write size (KB)} & 7.1  & 97.2 & 11.1  & 13.8 & 743.3 & 10.7  & 13.3 \\ \hline
\textbf{Workloads} & \textbf{Fuji-2} & \textbf{Tencent-0} & \textbf{Tencent-1} & \textbf{Tencent-2} & \textbf{Ali-0}  & \textbf{Ali-1}  & \textbf{Ali-2}  \\ \hline
\textbf{Read ratio (\%)}      & 87.6 & 84.3 & 2.0   & 98.2 & 98.1  & 81.3  & 11.0 \\ \hline
\textbf{Avg. read size (KB)}  & 39.3 & 31.2 & 12.5  & 47.0 & 37.0  & 370.4 & 26.0 \\ \hline
\textbf{Avg. write size (KB)} & 6.7  & 8.8  & 289.5 & 7.0  & 16.8  & 394.5 & 30.0 \\ \hline
\end{tabular}
}
\vspace{-5pt}
\caption{Workload characteristics.\label{tab:workload}}
\vspace{-22pt}
\end{table}

\begin{figure*}
    \begin{minipage}{0.76\textwidth}
        \includegraphics[width=1\linewidth]{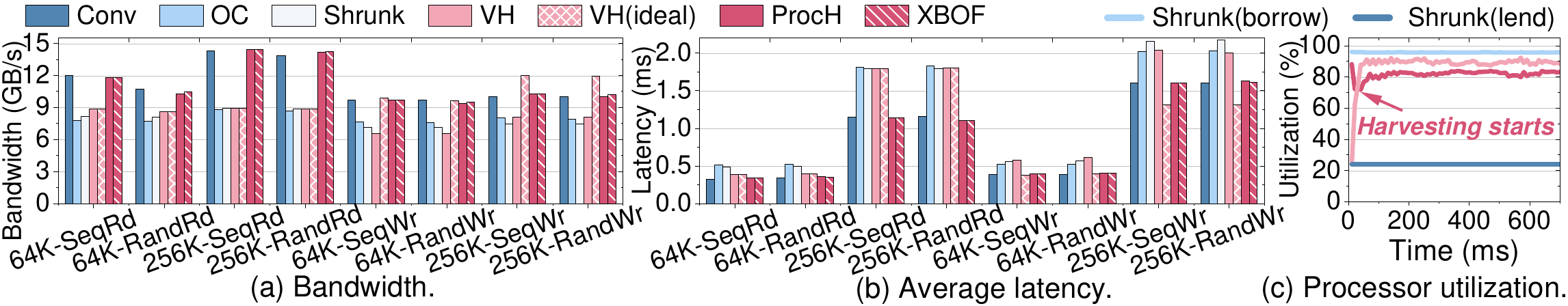}
        \vspace{-24pt}
        \caption{Performance benefits of processor harvesting. \label{fig:micro-proc}}
    \end{minipage}
    \begin{minipage}{0.235\textwidth}
        \includegraphics[width=1\linewidth]{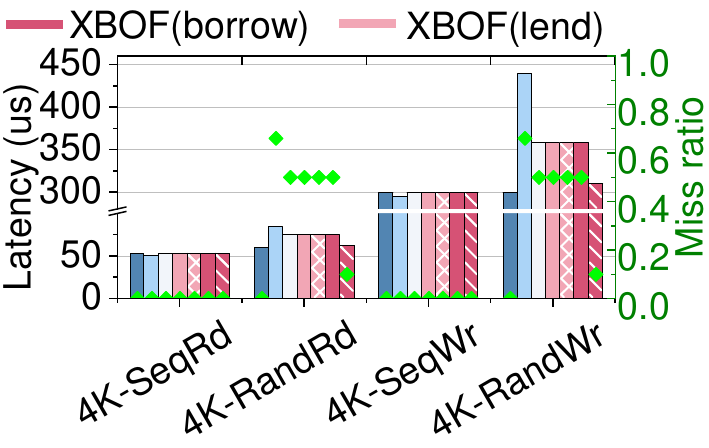}
        \vspace{-24pt}
        \caption{DRAM harvesting.\label{fig:micro-dram}}
    \end{minipage}
\vspace{-10pt}
\end{figure*}

\noindent \textbf{Workloads.}
We conduct evaluations with both microbenchmarks and various real workloads collected from production environments \cite{MSRC,kavalanekar2008characterization,yadgar2021ssd,systor,zhang2020osca,AlibabaBlockTrace}. 
Table \ref{tab:workload} lists their key characteristics.
To intuitively demystify the interactions between borrowers and lenders, we run workloads on 6 of the 12 SSDs (i.e., \texttt{borrowers}) by default, while keeping the other 6 SSDs idle (i.e., \texttt{lenders}). 
We also take sensitivity studies on varied numbers of borrowers and lenders in $\S$ \ref{sec:eval_sensitivity} and examine complex scenarios where all 12 SSDs have different workloads in $\S$ \ref{sec:eval_complex}. 
To demonstrate end-to-end performance improvement brought by our designs, we conduct comparisons on varied application instances in $\S$ \ref{sec:eval_numa}.

\subsection{Benefit Analysis}
\label{sec:eval_benefit}

\noindent \textbf{Processor harvesting.}
Figure \ref{fig:micro-proc} illustrates the throughput and average latency comparison in microbenchmarks. 
For simplicity, we present the average performance of the 6 SSDs that run workloads (i.e., \texttt{borrowers}).
We set the I/O depth as 64 to mimic the throughput-intensive scenarios and examine different I/O sizes from 64 KB to 256 KB. 
%
%
In comparison to \texttt{Conv}, \texttt{OC} and \texttt{Shrunk} suffer 27.8\% and 29.2\% throughput loss in all workloads, on average. 
Similar deterioration can also be observed in the latency comparison (i.e., 44.1\% and 46.4\% higher).
This is because the insufficient processor resources in \texttt{OC} and \texttt{Shrunk} cannot afford such intensive I/O patterns and become the performance bottleneck.
\texttt{VH} and \texttt{VH(ideal)} also lag far behind \texttt{Conv} in read workloads, as the simple virtualization and harvesting approach has no help to read requests (cf. $\S$ \ref{sec:ps_simple}).
Compared with \texttt{Shrunk}, \texttt{VH} and \texttt{VH(ideal)} succeed in improving write performance by redirecting parts of write requests to \texttt{lenders}.
VH(ideal) can even outperform \texttt{Conv} by 10.2\%. This can be attributed to the harvested flash resources of \texttt{lenders}, as data is temporarily stored in the \texttt{lenders'} flash backbone via their flash channels.
However, such gains are swept out after copyback occurs. 
As a result, \texttt{VH} still falls behind \texttt{Conv} by 25.6\%.
On the contrary, \texttt{XBOF} achieves comparable performance to \texttt{Conv} in all workloads with only half of the computing resources.
This is because \texttt{XBOF} can harvest idle processors of \texttt{lenders} for I/O serving.
Figure \ref{fig:micro-proc}\textcolor{fcolor}{c} shows the average processor utilization of \texttt{borrowers} and \texttt{lenders} in 256 KB sequential read test.
\texttt{XBOF} achieves 50.4\% higher utilization than \texttt{Shrunk}.

\noindent \textbf{DRAM harvesting.}
To evaluate how much XBOF can benefit from DRAM harvesting, we set an experiment to analyze the I/O performance in latency-sensitive scenarios (i.e., 4 KB reads and writes). We set the I/O depth to 1 in this test.
Figure \ref{fig:micro-dram} illustrates the average latency and mapping table miss ratio of different JBOF settings. 
%
%
Without sufficient DRAM to buffer the entire mapping table, \texttt{OC}, \texttt{Shrunk}, and \texttt{ProcH} experience 66.2\%, 49.7\%, and 49.7\% miss ratios in random read workloads, thereby causing 41.4\%, 24.7\%, and 24.7\% higher latencies, compared to \texttt{Conv}.
Similar degradation also exists in random write tests.
The simple virtualization and harvesting approach does not work for DRAM harvesting. 
This is because even if redirecting write requests to \texttt{lenders}, it still suffers from the miss penalty due to insufficient DRAM.
In contrast, the DRAM harvesting designs (cf. $\S$ \ref{sec:design_dram}) in \texttt{XBOF} enable \texttt{borrowers} to borrow idle DRAM resources from \texttt{lenders} to buffer their mapping table.
As a result, \texttt{XBOF} achieves comparable latencies to \texttt{Conv}.


\begin{figure}
    \centering
    \includegraphics[width=1\linewidth]{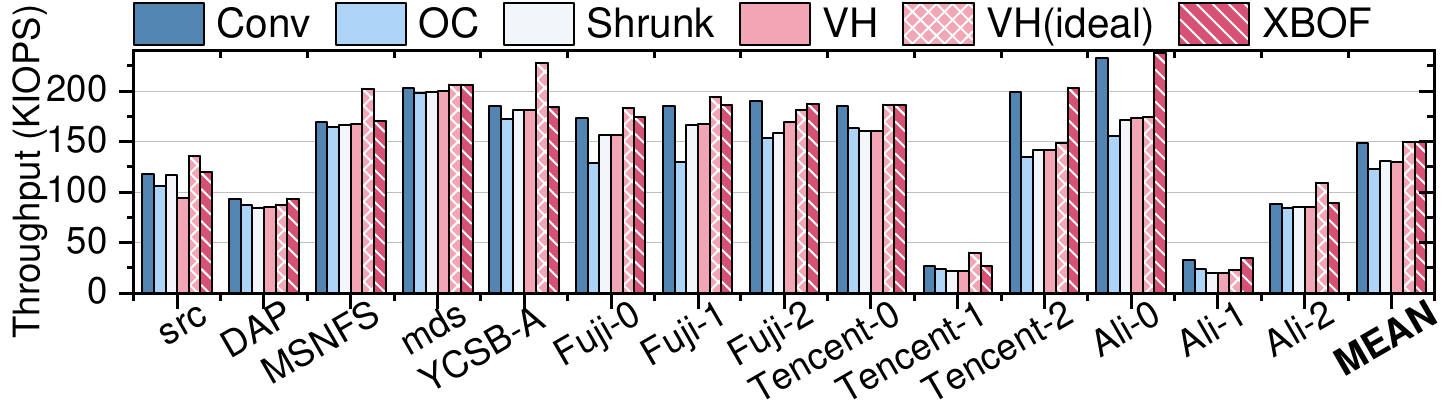}
    \vspace{-24pt}
    \caption{Throughput comparison in real workloads. \label{fig:macro}}
    \vspace{-10pt}
\end{figure}

\noindent \textbf{Improvements in real workloads.}
Figure \ref{fig:macro} presents the throughput comparison in diverse real workloads collected from production environments.
Compared with \texttt{Conv}, \texttt{OC} and \texttt{Shrunk} suffer 16.2\% and 13.4\% throughput loss in all workloads, on average, owing to the stressed computing resources.
\texttt{VH(ideal)} outperforms \texttt{Shrunk} in write-dominated workloads (e.g., 15.5\% in \texttt{src}) via write request redirecting. 
%
However, the substantial overhead of data copyback dispels such a mirage. Consequently, \texttt{VH} still lags behind \texttt{Conv} by 14.0\%. 
In contrast, \texttt{XBOF} can aid all workloads with diverse I/O types, eliminating the copyback overhead.
Specifically, although employing the same amount of computing resources, \texttt{XBOF} outperforms \texttt{Shrunk} and \texttt{VH} by 19.2\% and 20.0\%.
\texttt{XBOF} also achieves comparable throughput to \texttt{Conv}, which proves that our design can satisfy demanded performance targets via inter-SSD resource sharing.
%

\begin{figure}
    \centering
    \includegraphics[width=1\linewidth]{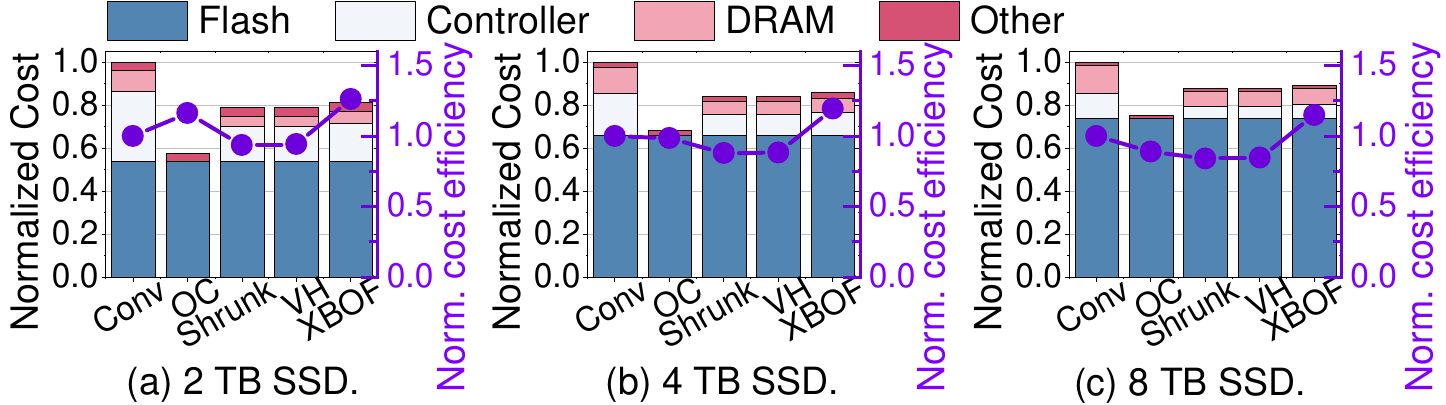}
    \vspace{-20pt}
    \caption{BOM cost analysis. \label{fig:cost}}
    \vspace{-10pt}
\end{figure}

\noindent \textbf{BOM cost saving.}
We further evaluate the BOM costs of SSDs in different JBOF platforms. 
According to the current prices on the market \cite{ChinaFlashMarket,dramexchange,microndram,TrendForce,Maxio,Longsys,ssdcalculator}, we identify the costs of NAND flash, DDR4 DRAM, enterprise SSD controller, and other expenses (e.g., PCB board and packaging) as \$4.95 per 128 GB, \$7.2 per GB, \$48, and \$6, respectively. 
We assume the halved computing resources (i.e., SSD controller and DRAM) in \texttt{Shrunk} and \texttt{VH} consume halved costs. According to prior work \cite{tang2024exploring}, we estimate the prices of CXL-enabled SSD controller and DRAM in \texttt{XBOF} are 10\% higher than those in \texttt{Shrunk}.
As shown in Figure \ref{fig:cost}, \texttt{XBOF} succeeds in saving BOM cost by 19.0\% for 2 TB SSDs, compared with \texttt{Conv}.
Although SSDs in \texttt{XBOF} are more expensive than those in \texttt{OC} and \texttt{Shrunk}, such expenses are worth given the improved performance. 
Figure \ref{fig:cost} also depicts the cost efficiency in \texttt{Ali-0} workload.
We define cost efficiency as the bandwidth achieved by per unit cost. 
\texttt{XBOF} outperforms all other designs (e.g., 19.7\% higher than \texttt{OC}) in this metric.

\subsection{Overhead Analysis}
\label{sec:eval_overhead}

\begin{figure}
    \centering
    \includegraphics[width=1\linewidth]{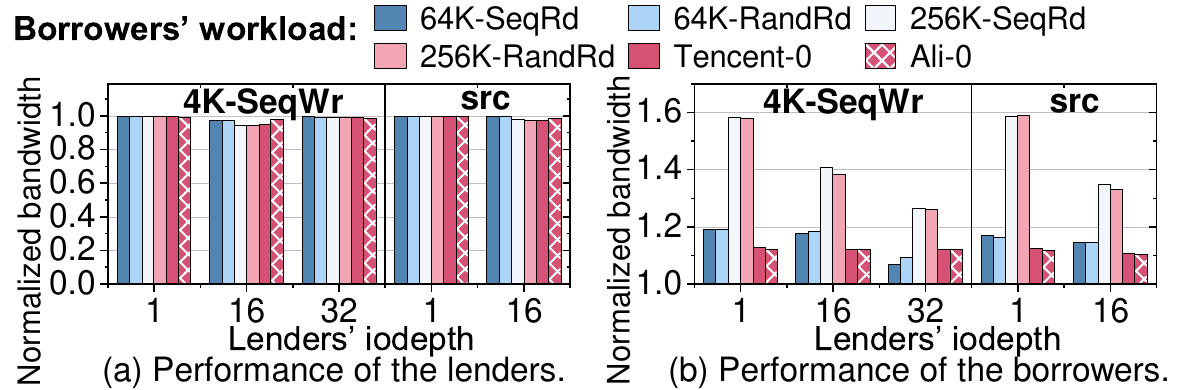}
    \vspace{-20pt}
    \caption{Interaction between lenders and borrowers. \label{fig:overhead-lender}}
    \vspace{-10pt}
\end{figure}

\noindent \textbf{Performance impact on lenders.}
In this test, we run varied I/O-intensive workloads on \texttt{borrowers}, while \texttt{lenders} serve moderate I/O requests.
We set the I/O depth as 64 for \texttt{borrowers} while varying the I/O depth from 1 to 32 for \texttt{lenders} to mimic different degrees of I/O pressure.
Note that \texttt{lenders'} processors are too busy to be lent when running \texttt{src} workload in 32 I/O depth. Therefore, we omit this result.
Figure \ref{fig:overhead-lender} presents the throughput of \texttt{lenders} and \texttt{borrowers} in \texttt{XBOF}.
We have normalized the results to that of \texttt{Shrunk}, where no resource is lent out (i.e., best case for \texttt{lenders}) or borrowed in (i.e., worst case for \texttt{borrowers}).
Resource lending causes negligible performance loss (1.3\% on average, cf. Figure \ref{fig:overhead-lender}\textcolor{fcolor}{a}) for \texttt{lenders}. This is because, with our holistic load balance algorithm (cf. $\S$ \ref{sec:design_processor}), \texttt{lenders} can reserve sufficient computing resources to handle their own I/O commands.
The throughput of \texttt{borrowers} improves 15.5\%, 23.3\%, and 30.0\%, in the tests where the \texttt{lenders} serve 4 KB sequential writes in I/O depths of 32, 16, and 1. 
This is because, with lighter workloads, \texttt{lenders} can lend out more resources to help \texttt{borrowers} serve I/O commands. 

\begin{figure}
    \centering
    \includegraphics[width=1\linewidth]{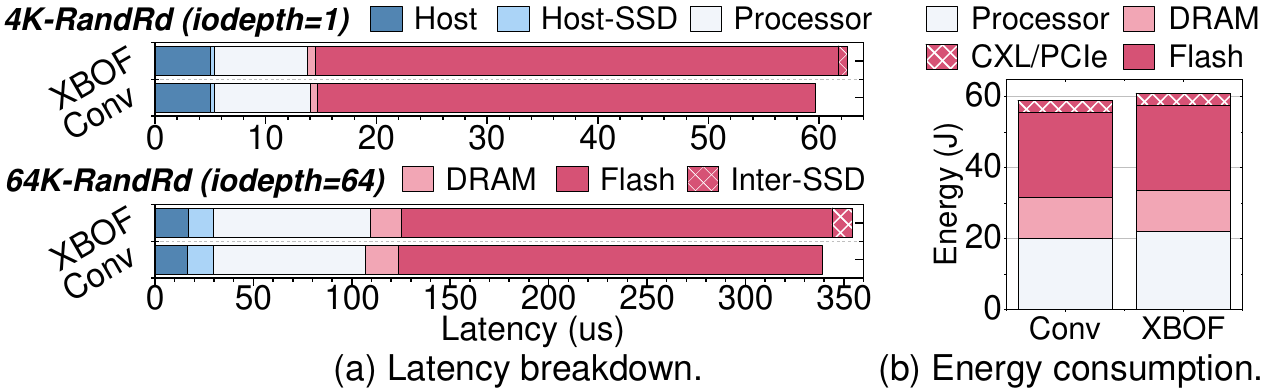}
    \vspace{-20pt}
    \caption{Overhead analysis (latency and energy). \label{fig:overhead-lat-energy}}
    \vspace{-10pt}
\end{figure}

\noindent \textbf{Extra latency.}
XBOF designs (e.g., remote metadata access and synchronization) can introduce extra latency.
To understand such overhead, we break down the latency of 4 KB and 64 KB random reads into six parts, as shown in Figure \ref{fig:overhead-lat-energy}\textcolor{fcolor}{a}. 
\texttt {Host} is the time consumed by the host I/O stack (e.g., NVMe driver).
\texttt{Host-SSD} is the time for data and NVMe command transfer between the host and SSD.
\texttt{Processor} is the time consumed to execute SSD firmware (e.g., I/O parsing and address translation).
\texttt{DRAM} encloses the time of onboard DRAM accesses (e.g., reading mapping table).
\texttt{Flash} represents the time of flash operations (e.g., flash read).
Finally, \texttt{Inter-SSD} includes the time taken on the CXL interconnection.
%
In both \texttt{Conv} and \texttt{XBOF}, \texttt{Flash} dominates the latency. 
Compared with \texttt{Conv}, \texttt{XBOF} causes 3.3\% more \texttt{Flash} overhead because of the sporadic DRAM miss (cf. $\S$ \ref{sec:eval_benefit}).
\texttt{XBOF} also takes 3.1\% more \texttt{Processor} time for inter-SSD synchronization.
There is no obvious difference in terms of \texttt{Host} overhead, thanks to our decentralized resource management scheme (cf. $\S$ \ref{sec:design_management}).
Moreover, XBOF only takes 20 ns more host CPU time for each I/O command to
compute the load balance formula for redirecting (cf. $\S$ \ref{sec:design_processor}).
%
\texttt{XBOF} takes minor \texttt{Inter-SSD} time (up to 2.9\%), because CXL interconnection has extremely high speed and delivers sub-microsecond remote access. 

\noindent \textbf{Energy consumption.}
Figure \ref{fig:overhead-lat-energy}\textcolor{fcolor}{b} illustrates the energy consumption to conduct \texttt{Fuji-0} workload of different designs. 
Compared with \texttt{Conv}, \texttt{XBOF} takes 3.5\% more energy, because of the added XBOF daemon and CXL-enabled inter-SSD communication (cf. $\S$ \ref{sec:design_ssd}).
However, this minor overhead brings huge rewards, as it allows 
resource sharing to exploit the idle SSDs in JBOF, thereby achieving required I/O performance with reduced resources and costs (cf. $\S$ \ref{sec:eval_benefit}).

\begin{figure}
    \centering
    \includegraphics[width=1\linewidth]{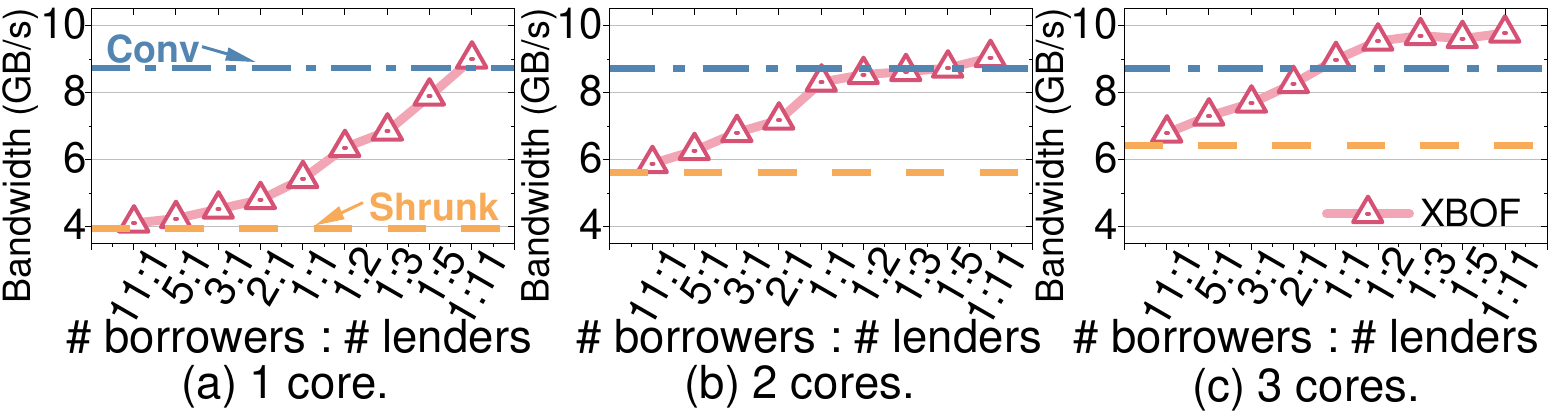}
    \vspace{-23pt}
    \caption{Sensitivity study on different processor resources. \label{fig:sensitivity-processor}}
    \vspace{-25pt}
\end{figure}

\begin{figure*}
    \begin{minipage}{0.39\textwidth}
        \includegraphics[width=1\linewidth]{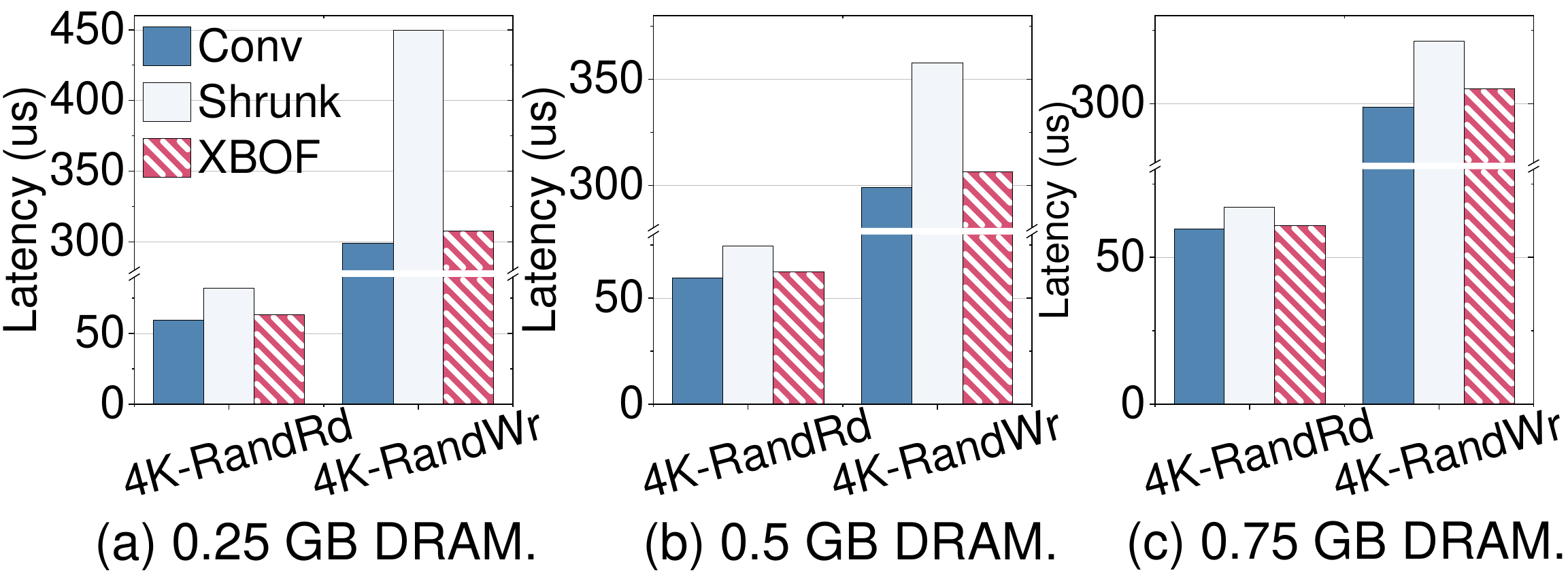}
        \vspace{-23pt}
        \caption{Analysis of varied DRAM resources. \label{fig:sensitivity-dram}}
    \end{minipage}
    \begin{minipage}{0.39\textwidth}
        \includegraphics[width=1\linewidth]{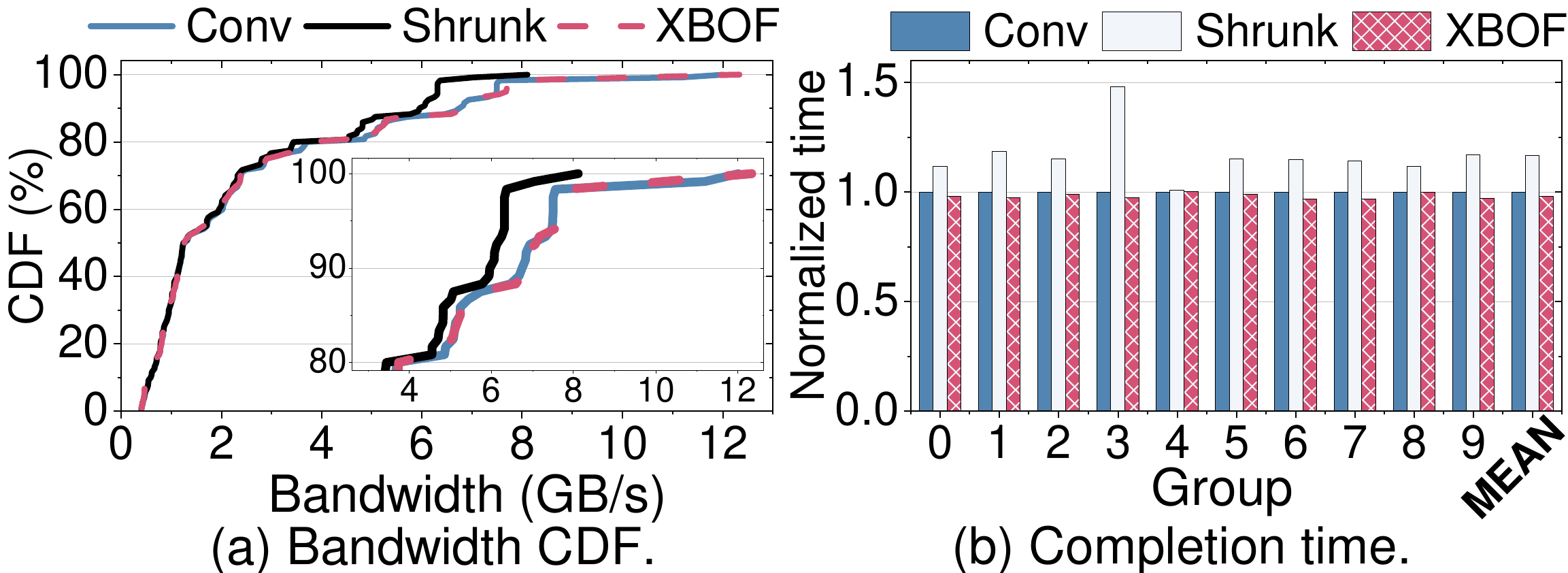}
        \vspace{-23pt}
        \caption{Comparison in complex scenarios.\label{fig:complex}}
    \end{minipage}
    \begin{minipage}{0.2\textwidth}
        \includegraphics[width=1\linewidth]{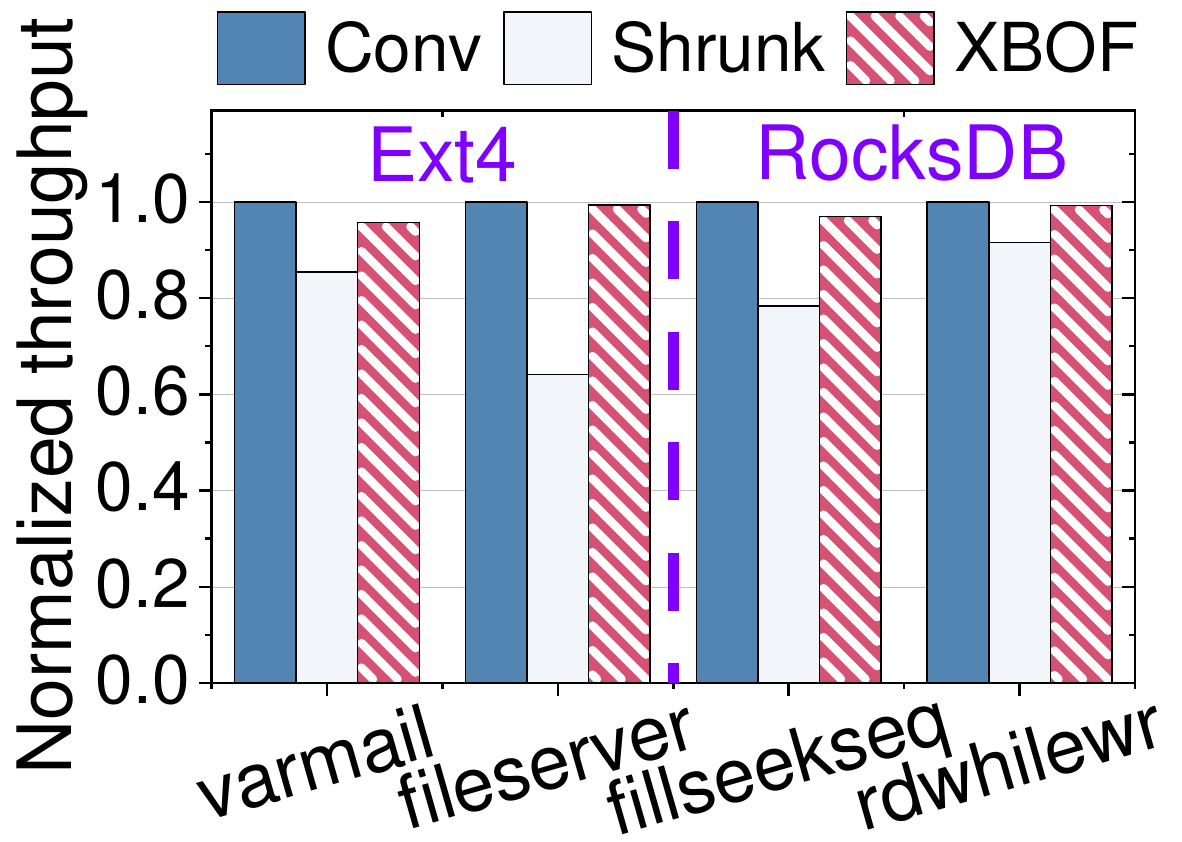}
        \vspace{-23pt}
        \caption{NUMA test.\label{fig:numa}}
    \end{minipage}
\vspace{-6pt}
\end{figure*}

\subsection{Sensitivity Study}
\label{sec:eval_sensitivity}

\noindent \textbf{Different processor resources.}
We examine the benefits brought by our designs in different processor resource configurations. 
For fair comparisons, we equip SSDs in \texttt{Shrunk} and \texttt{XBOF} with the same capacity of DRAM as SSDs in \texttt{Conv}.
We vary the number of ARM cores in each SSD of \texttt{Shrunk} and \texttt{XBOF} from 1 to 3.
We also change the ratios of numbers of \texttt{borrowers} and \texttt{lenders} from 11:1 to 1:11. 
Figure \ref{fig:sensitivity-processor} shows the variation of throughput in \texttt{Ali-0} workload.
Without inter-SSD resource harvesting, the throughput of \texttt{Shrunk} decreases with the decreasing number of cores (up to 54.6\% degradation in the 1-core setting, cf. Figure \ref{fig:sensitivity-processor}\textcolor{fcolor}{a}).
In contrast, \texttt{XBOF} achieves comparable throughput to \texttt{Conv} when there are enough \texttt{lenders} for harvesting.
For example, in 2-core tests, \texttt{XBOF} achieves 97.7\% performance of \texttt{Conv} when each \texttt{borrower} harvests two \texttt{lenders} (i.e., \texttt{1:2}).
%
Note that, with excessive \texttt{lenders} for harvesting (e.g., when ratio is \texttt{1:11}), \texttt{XBOF} cannot further boost performance, owing to the limited throughput of flash backbone and overwhelming synchronization overhead.

\noindent \textbf{Different DRAM resources.}
We reserve different capacities of DRAM in the SSDs within \texttt{Shrunk} and \texttt{XBOF}.
For fair comparisons, we equip SSDs in all JBOF platforms with 6 ARM cores.
We assume there is enough number of idle SSDs for DRAM harvesting, which has been witnessed in production environments (cf. $\S$ \ref{sec:bg_motivation}). 
Figure \ref{fig:sensitivity-dram} illustrates the latency comparison. 
%
\texttt{Shrunk} experiences 44.0\%, 22.3\%, and 10.0\% higher latency with 0.25, 0.5, and 0.75 GB DRAM per TB flash capacity, respectively.
On the contrary, benefiting from our DRAM harvesting designs (cf. $\S$ \ref{sec:design_dram}), \texttt{XBOF} only introduces negligible latency increases (3.4\% on average).

\subsection{Complex Scenario}
\label{sec:eval_complex}

We examine complex scenarios where all SSDs have their own workloads.
We randomly select 12 workloads from Tencent \cite{zhang2020osca} traces as a group, and assign each workload to a single SSD. We repeat the experiment 10 times.
%
Figure \ref{fig:complex}\textcolor{fcolor}{a} presents the cumulative distribution function (CDF) of throughput across all 120 workloads. 
Echoing our findings in the previous tests, \texttt{XBOF} succeeds in fulfilling the burst I/O performance demands, even with only halved computing resources. 
To be specific, SSDs in \texttt{XBOF} achieve 12.3 GB/s peak throughput, while this value is only 8.1 GB/s in \texttt{Shrunk}.
Figure \ref{fig:complex}\textcolor{fcolor}{b} shows the workload completion time comparison in different groups.
Compared with \texttt{Shrunk}, \texttt{XBOF} shortens at most 34.3\% completion time (15.2\% on average), thanks to our inter-SSD resource sharing designs. 

\subsection{End-to-end Application on NUMA Platform}
\label{sec:eval_numa}
We now evaluate XBOF designs on a 2-socket NUMA-based emulation platform (cf. $\S$ \ref{sec:design_imp}). We use one socket to mimic the borrower SSD, while the other socket acts as the lender SSD and the host.
We choose Ext4 filesystem \cite{mathur2007new} and RocksDB \cite{rocksdb} as the representative applications of JBOF and test them with \texttt{filebench} \cite{filebench} and \texttt{db\_bench} \cite{db_bench}, respectively.
Figure \ref{fig:numa} shows the throughput comparison of different designs.
Similar to our previous tests, \texttt{XBOF} outperforms \texttt{Shrunk} by 24.8\% and achieves comparable throughput to \texttt{Conv} even with reduced computing resources, cross-validating the superiority of our designs.

\section{Related Work and Discussion}
\label{sec:relatedwork}
\noindent \textbf{SSD architecture.}
Multiple studies \cite{kim2022networked,nadig2023venice,kim2023decoupled,jung2022hello,yang2023overcoming,zhang2020scalable,peng2025xharvest} have been proposed to renovate the SSD architectures.
Decoupled SSD \cite{kim2023decoupled} decomposes SSD architecture into front-end (i.e., SSD controller) and back-end (i.e., flash backbone) and introduces a network-on-chip to facilitate communication among flash controllers. It improves the efficiency of SSD internal data movement (i.e., garbage collection).
In contrast, XBOF disaggregates SSD components based on their functionalities to enable fine-grained resource management and sharing.
%
XHarvest \cite{peng2025xharvest} renovates SSD architecture and achieves high I/O performance through secure host resource harvesting. On the contrary, XBOF satisfies performance requirement via inter-SSD resource sharing.
%

\noindent \textbf{Communication protocol.}
NVMe is the de-facto communication protocol for most high-performance SSDs. 
We opt to implement I/O redirection and load balance on NVMe protocol to be compatible with existing I/O stacks, avoiding excessive software modification. 
To adapt to prior works \cite{jung2022hello,kwon2023cache,yang2023overcoming,li2025bytefs, zhang2025skybyte} that access SSDs via \texttt{load/store} instructions, a similar I/O redirection and load balance mechanism can be implemented in the memory access path (e.g., the memory management subsystem of Linux kernel \cite{linux_memory}).  

\noindent \textbf{Storage virtualization.}
Great efforts \cite{reidys2022blockflex,min2021gimbal,yang2018spdk,huang2017flashblox,kwon2020fvm,russell2008virtio,sun2025fleetio} have been taken to virtualize storage devices and improve their utilization.
BlockFlex \cite{reidys2022blockflex} and FleetIO \cite{sun2025fleetio} enhance SSD utilization by harvesting idle flash resources, thereby facing critical challenges, such as limited read profit and huge data copyback overhead (cf. $\S$ \ref{sec:ps_simple}).
On the contrary, XBOF focuses on the stateless computing resources and enables general and lightweight inter-SSD resources sharing with the support of CXL interconnection.


\noindent \textbf{RAID.}
RAID \cite{mdraid,yi2022scalaraid,chen1994raid,jiang2021fusionraid,kim2023raizn,yi2024biza,shu2023disaggregated} is a storage organization scheme that can balance I/O requests among multiple SSDs, thereby improving SSD utilization. This technique is orthogonal to XBOF.
Specifically, users typically construct RAID with SSDs from different JBOFs in different racks \cite{shu2023disaggregated} for higher fault tolerance.
These RAIDs serve varied applications with diverse utilization patterns, which deliver opportunities for inter-RAID resource harvesting.

\noindent \textbf{JBOF with heterogeneous SSDs.}
Our design can be adapted to JBOFs comprised of heterogeneous SSDs with minor revisions. 
Specifically, the borrower can expose its firmware tasks as executable files \cite{yang2023lambda,larus1994rewriting} in trusted execution environments (TEE) \cite{bauman2018sgxelide,kang2021iceclave,peng2025xharvest}. Therefore, the lender can seamlessly execute the borrower's firmware tasks with its general-purpose ARM processor, while ensuring security. 
Considering the computing power disparity of varied SSDs, XBOF can replace the busy indicator (which is processor utilization now, cf. $\S$ \ref{sec:design_processor}) with a more general and absolute metric (e.g., current waiting queue depth) for load balance.

\noindent \textbf{Practicality and compatibility.}
The main hardware modifications of XBOF are only the CXL interconnections and reduced SSD computing resources, while the other innovations (e.g., XBOF daemon) are implemented as software/firmware.
Thanks to the backward-compatibility of CXL to PCIe and our support to NVMe protocol, XBOF SSD can act as a conventional SSD and adapt to existing I/O stacks and applications without violating compatibility.


\section{Conclusion}
\label{sec:conclusion}
While the numerous computing resources significantly elevate the BOM costs of SSDs to satisfy the performance requirement of burst I/O, the sporadic nature of I/O bursts causes severe SSD underutilization in JBOF scenarios.
Tackling this issue, we propose XBOF, a novel JBOF design that reserves moderate computing resources in SSDs at low costs while achieving demanded I/O performance by employing CXL to facilitate fine-grained inter-SSD resource sharing.

\bibliographystyle{ACM-Reference-Format}
\bibliography{ref}

\end{sloppypar}
\end{document}